\documentclass[%
 reprint,
 amsmath,amssymb,
 aps,
]{revtex4-2}

\usepackage{graphicx}
\usepackage{dcolumn}
\usepackage{bm}

\begin{document}

\preprint{APS/123-QED}

\title{Geometry-induced friction at a soft interface}

\author{Aashna Chawla}
\author{Deepak Kumar}%
 \email{krdeepak@physics.iitd.ac.in}
\affiliation{Department of Physics, Indian Institute of Technology Delhi, Hauz Khas, New Delhi, India
}%





\begin{abstract}
Soft and biological matter come in a variety of shapes and geometries. When soft surfaces that do not fit into each other due to a mismatch in Gaussian curvatures form an interface, beautiful geometry-induced patterns emerge \cite{Hure2012,Hure2011,King2012,tobasco2022exact}. In this paper, we study the effect of geometry on the dynamical response of soft surfaces moving relative to each other. Using a novel experimental scheme, we measure friction between a highly bendable thin polymer sheet and a hydrogel substrate. At this soft and low-friction interface \cite{gong1997friction,Cuccia2020}, we find a strong dependence of friction on the relative geometry of the two surfaces -  a flat sheet experiences significantly larger friction on a spherical substrate than on planar or cylindrical substrate. We show that the stress developed in the sheet due to its geometrically incompatible confinement \cite{Hohlfeld2015,Hure2011} is responsible for the enhanced friction. This mechanism also leads to a transition in the nature of friction as the sheet radius is increased beyond a critical value. Our finding reveals a hitherto unnoticed non-specific mechanism of purely geometrical origin that may influence friction significantly in soft, biological, and nano-scale systems. In particular, it provokes us to re-examine our understanding of phenomena such as the curvature dependence of biological cell mobility \cite{Pieuchot2018}.
\end{abstract}

\maketitle


The coming together of soft objects with incompatible geometries often leads to a rich glossary of patterns and phenomena, for example, the plethora of phases seen in bent-core liquid crystals \cite{2006JaJAP..45..597T}, size-selection in the self-assembly of systems with incompatible building blocks \cite{Niv2018, Lenz2017} and the emergence of beautiful wrinkle patterns when thin sheets are confined to substrates with incompatible geometries \cite{Hure2012,Hure2011,King2012,tobasco2022exact}. In the last case above, the incompatibility lies in the Gaussian curvature mismatch between the thin sheet and the substrate and is a consequence of the Gauss's Theorema Egregium \cite{Struik1988}. This problem has been studied in various settings, including a planar sheet on spherical liquid drop \cite{King2012}, a planar sheet on a spherical solid substrate \cite{Hohlfeld2015}, and a thin spherical shell on a planar liquid surface \cite{tobasco2022exact}. The ground state obtained in these problems usually involves a non-trivial stress distribution arising due to geometrical incompatibility. While many recent papers have studied the effect of such geometrical incompatibility-induced stress distribution on \emph{static} wrinkle patterns, its effect on the dynamics of such systems remains relatively unexplored.

In this paper, we study the effect of geometrical incompatibility on friction at a soft interface subject to small relative velocity ($v\sim 10 nm/s$). Friction is an intrinsic feature associated with relative motion between two bodies. 
Given its ubiquity and importance, friction has been studied since very early times, 
yet many fundamental questions remain open \cite{urbakh2004nonlinear,vanossi2013colloquium}. Particularly, recent advances in nano- and bio-tribology have brought many surprises, for example, the recent observation of the dependence of friction on relative geometry and commensurability of two-dimensional layered atomic materials sliding relative to each other \cite{Vanossi2020,superlubricity}.

The interface studied in this paper consists of a thin planar elastic sheet placed on a low-friction hydrogel substrate having different geometrical shapes, viz. planar, cylindrical, and spherical. We find that the frictional force at this interface shows a strong geometry dependence and is significantly larger for the geometrically incompatible configuration of a planar sheet-spherical substrate than the other two geometrically compatible configurations: planar sheet-planar substrate and planar sheet-cylindrical substrate. We also observe a dependence of friction on the sheet size and see a transition in the friction behavior at an intermediate value of sheet radius for the flat sheet-spherical substrate system. We show that these effects are a consequence of the coupling of the stress developed in the sheet due to its geometrically incompatible confinement with the curvature of the interface, resulting in an increased normal force.

\begin{figure*}
    \begin{center}
        \includegraphics[width=0.95\textwidth]{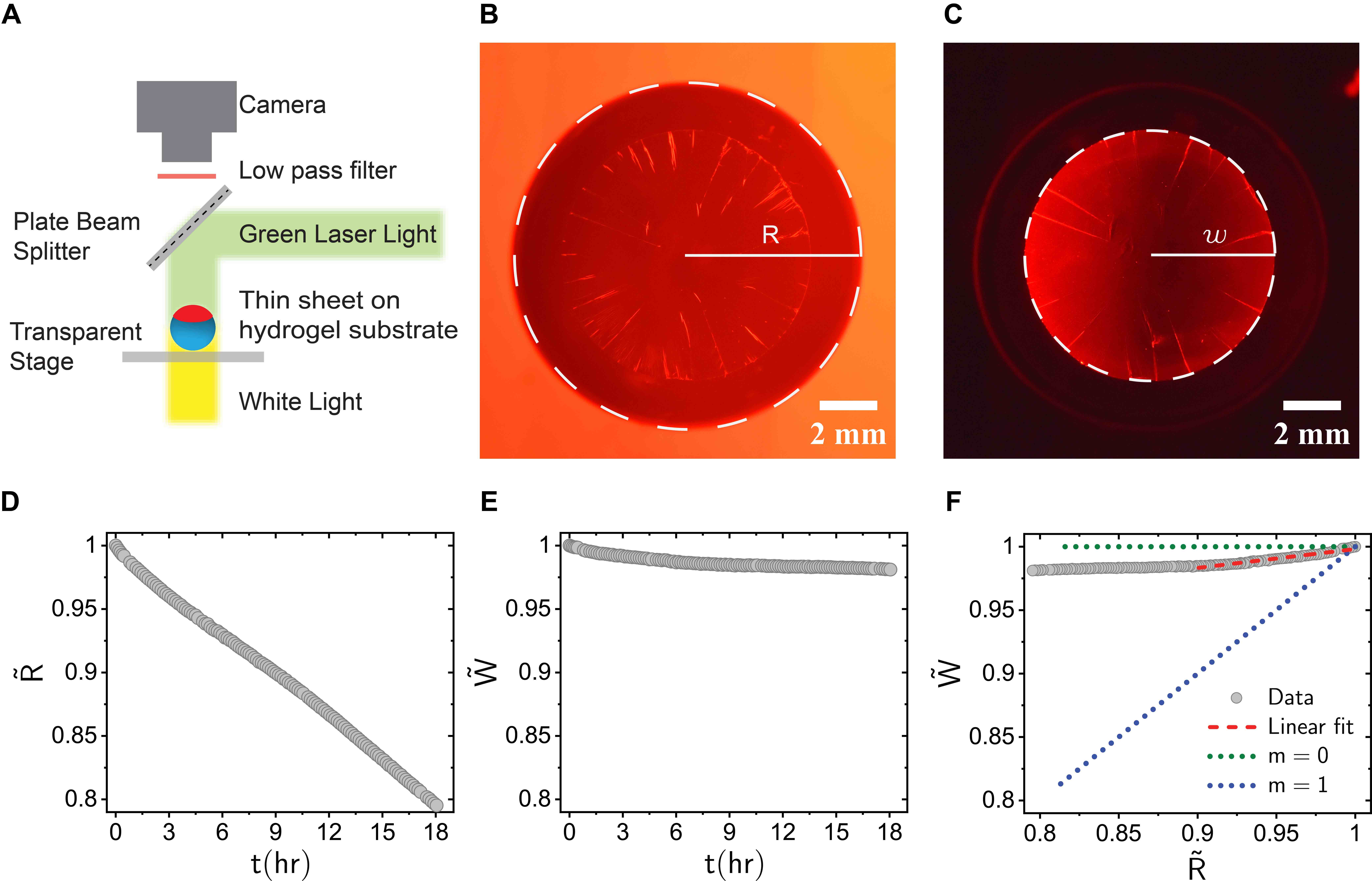}
        
    \end{center}
    \caption{\textbf{Geometry dependent tribometery.} (A) Schematic of the experimental setup. (B) A typical image of the spherical hydrogel substrate obtained with bright field transmission mode imaging. The image is analyzed to obtain the substrate radius $R$. (C) Typical image of the sheet captured by fluorescence imaging, which is used to obtain the projected radius $w$ of the sheet. (D) The variation of the normalized substrate radius with time $\Tilde{R}(t)$. (E) The variation of the normalized sheet radius with time $\Tilde{W}(t)$. (F) $\Tilde{W}$ plotted as a function of $\Tilde{R}$. A linear fit to the initial part of the function $\tilde{W}(\tilde{R}$) (red dashed line) is used to obtain the slope $m$. The limits $m=0$ (green dashed line) and $m=1$ (blue dashed line) represent no friction and large friction, respectively. The value of $m$ is a measure of the strength of friction.}
   \label{fig:Figure1}
\end{figure*}

The experimental setup is shown schematically in Fig. \ref{fig:Figure1}A. Thin planar sheets of polystyrene (thickness $h = $ $50\: nm $ to $ 800 \:nm $) 
cut into circular discs of radius $W_0$ are placed on the surface of swollen hydrogel substrates. The hydrogel substrate loses water through evaporation in the ambient maintained at a temperature of $24^\circ C$, and shrinks in size, generating relative motion at the thin sheet-hydrogel interface. We monitor the relative change between the sheet and substrate sizes and use it to obtain a measure of friction similar to the method of measuring the viscosity of a liquid from the terminal velocity of a sedimenting sphere.

We image the substrate and the sheet separately using two different imaging schemes implemented in sequence: a brightfield transmission mode imaging for the substrate and fluorescence imaging for the sheet which is tagged with the fluorescent dye Nile red. Typical images of the substrate (in this case, a sphere) and the sheet are shown in Fig. \ref{fig:Figure1}B,C, respectively. Pairs of such images, recorded at equal intervals of time ($\Delta t \sim 10 \:minutes$), are analyzed to obtain the substrate radius ($R$) and the sheet radius ($w$). 
The projected radius of the sheet measured from the image ($w$) is used to obtain the real radius of the sheet measured along the curved surface of the substrate: $W=R sin^{-1}(w/R)$.

We calculate the normalized substrate and sheet radii at any instant of time $t$ as $\Tilde{R}(t)=R(t)/R_0$ and $\Tilde{W}(t)=W(t)/W_0$ where $R_0$ and $W_0$ are their initial radii, 
respectively. 
We observe that $\Tilde{R}(t)$ usually decreases at a constant rate (Fig. \ref{fig:Figure1}D where $\dot{\Tilde{R}}=-0.011 hr^{-1}$), as expected for evaporation-driven water loss proportional to the surface area. A decrease in $\Tilde{R}$ is accompanied by a corresponding decrease in $\Tilde{W}$ (Fig. \ref{fig:Figure1}E), 
due to the force of friction that opposes any relative motion between the sheet and the substrate. If the sheet-substrate interface was friction-less, the radius of the sheet would not change at all with time, and there would be large relative motion at the interface. On the other hand, in the limit of large friction, there would be no relative motion between the sheet and the substrate, and the radius of the sheet would change by the same factor as the substrate. 
On the $\Tilde{R}-\Tilde{W}$ plot (Fig. \ref{fig:Figure1}F), these two limiting cases correspond to straight lines with slopes $m=0$ and $1$, respectively. The above discussion suggests that $m$ can be used as a measure of the strength of the frictional interaction. 
We show in SI that the velocity dependence of friction for hydrogels \cite{Simič2020,gong1997friction,wu2021relaxation,Cuccia2020}
allows us to write friction in terms of $m$ and that for small values of $m$, friction is proportional $m$. 
Our setup thus gives us a way to determine the small force of friction between the sheet and the substrate, thus working like a sensitive tribometer.

The $\tilde{W}(t)$ and $\tilde{R}(t)$ data presented in Fig. \ref{fig:Figure1}D,E respectively, when plotted as $\tilde{W}(\tilde{R})$ lies between the two straight lines corresponding to 
$m=0$ and $1$. When considered over a large range of $\tilde{R}$, $\tilde{W}(\tilde{R})$ is logarithmic, 
a behavior reminiscent of the logarithmic relaxation observed in experiments on crumpling of thin sheets \cite{Matan2002, Lomholt2013} (see SI for details). 
For the present discussion, we restrict ourselves to a small initial part $1>\Tilde{R}\gtrsim 0.9$ of the $\Tilde{W}(\Tilde{R})$ curve and fit a straight line to obtain the slope $m$.

\begin{figure*}
    \begin{center}
    
      \includegraphics[scale=1]{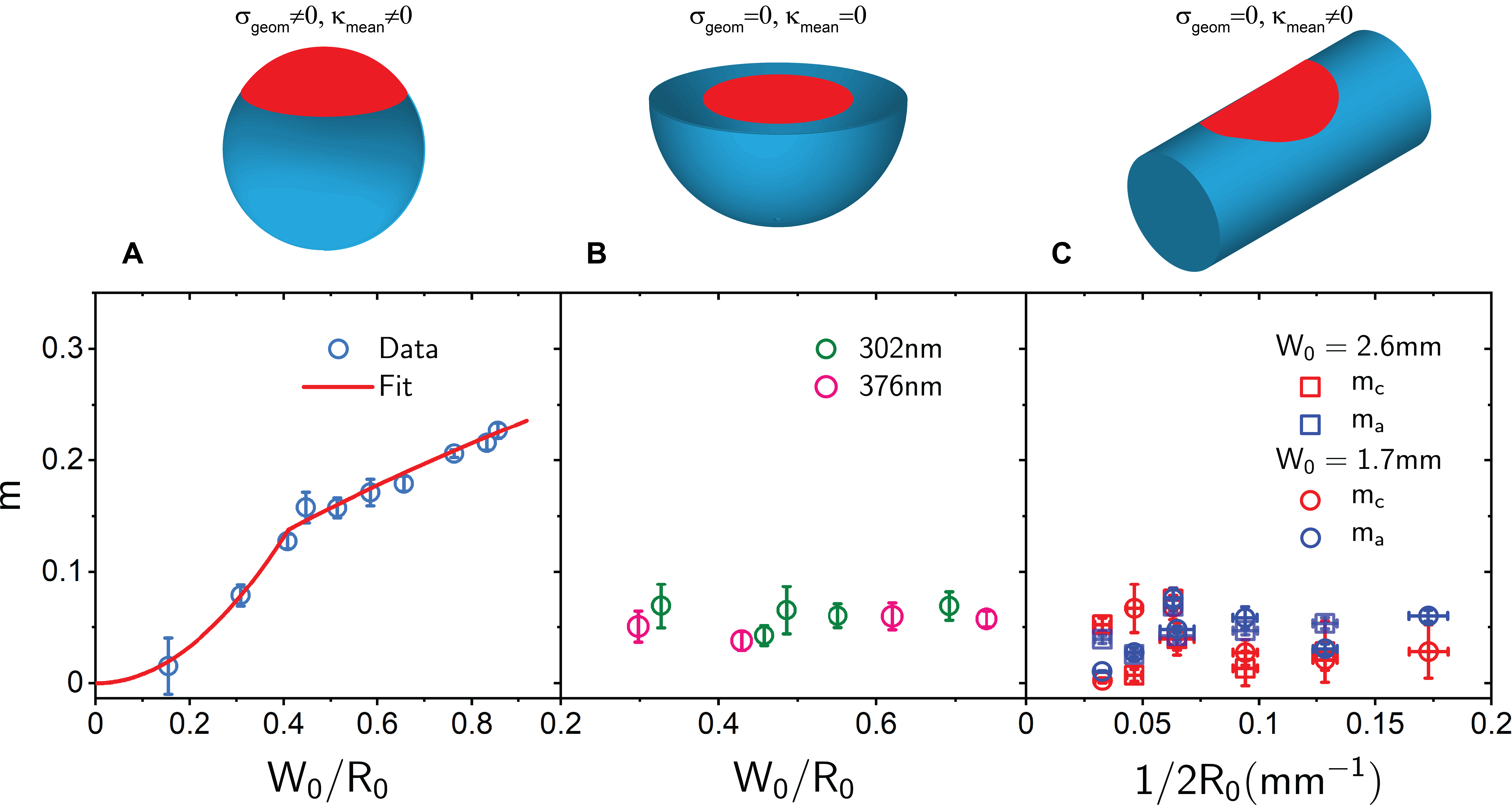}  
    \end{center}
    
    \caption {\textbf{Effect of geometrical incompatibility on friction.}
    (A) $m(W_0/R_0)$ for flat sheets ($h=302 nm$) on spherical substrates ( $R_0\sim 6mm$). 
    The red curve is a fit to equation (\ref{eq:slope}). 
    (B) $m(W_0/R_0)$ for flat sheets ($h=302nm$, $376nm$) on flat substrates. 
    (C) Variation of the slopes measured along the axial direction $m_a$ (blue) and the curved direction $m_c$ (red) with the mean curvature $1/2R_0$ for flat sheets ($h=376nm$, $W_0=2.6 mm$ and $1.7 mm$) on cylindrical substrates. We observe larger values of $m$ in (A) than in (B) or (C). (B) and (C) are geometrically compatible configurations, as the Gaussian curvatures of the sheet and the substrate match, and therefore $\sigma_{geom}=0$. On the other hand, (A) is an incompatible configuration with $\sigma_{geom}\neq0$ and a corresponding pressure $\sigma_{geom}.\kappa_{mean}$. } 
    

    \label{fig:Figure2}
\end{figure*}

As with most materials, friction in hydrogels is proportional to the normal force; however, with a very small coefficient of friction ($\sim 0.01$) \cite{Cuccia2020,gong2006friction}. Gravitational force, which is the dominant contributor to the normal force for macroscopic objects, has a very small value for the thin sheets ($\sim \rho g W^2 h$). However, for the data in Fig. \ref{fig:Figure1}F, we obtain a finite value of $m=0.15$. When we repeat the experiment by placing sheets of different initial radii $W_0$, on spherical substrates of initial radius $R_0\approx 6mm$, we observe that the value of $m$ increases monotonically with $W_0/R_0$ as shown in Fig. \ref{fig:Figure2}A. We note that in the flat sheet-spherical substrate geometry, the strength of geometrical confinement depends on the parameter $W_0/R_0$. For a small value of $W_0/R_0$, the sheet sees almost a flat substrate, while for a value of $W_0/R_0 \sim 1$, the sheet feels the curvature of the substrate strongly. Therefore, the dependence of $m$ on $W_0/R_0$ indicates that the force of friction is being affected by the substrate geometry.

In order to further verify 
the role of geometry, we measure $m$ for a flat sheet on a flat hydrogel substrate. 
Figure \ref{fig:Figure2}B shows $m(W_0/R_0)$ for $h=302nm$ and $376nm$, with $R_0\approx 6mm$. We see that a flat sheet 
``slips'' on a flat substrate, with 
$m<0.1$. The larger value of $m$ for the flat sheet-spherical substrate than for the flat sheet-flat substrate configuration confirms the role of curvature. However, a flat surface has neither Gaussian curvature ($\kappa_{Gauss}=0$) nor mean curvature ($\kappa_{mean}=0$), while a sphere has 
both: $\kappa_{Gauss}=1/R^2$ and $\kappa_{mean}=1/R$. With an aim to untangle the effects of the mean and Gaussian curvatures, we perform experiments on cylindrical substrates having $\kappa_{mean}=1/2R$ and $\kappa_{Gauss}=0$. 

In our experiments on cylindrical substrates of radius $R$ and axial length $M$, we measure the sheet radii along both the axial direction ($W_a$) and the curved direction ($W_c$) and normalize them with their initial values, respectively: $\tilde{W_a}=W_a/W_a(t=0)$, and, $\tilde{W_c}=W_c/W_c(t=0)$. We then determine the slopes of the $\tilde{W_a}(\tilde{M})$ and $\tilde{W_c}(\tilde{R})$ curves to obtain $m_a$ and $m_c$, respectively. Figure \ref{fig:Figure2}C shows the variation of $m_a$ and $m_c$ with the initial mean curvature of the cylindrical substrate $1/{2R_0}$. 
We observe that both $m_{a}$ and $m_c$ have small values ($<0.1$), with no clear trend of variation with increasing mean curvature.  We, therefore, conclude that the substrate's Gaussian curvature plays an important role in the mechanism responsible for the larger value of $m$ in flat sheet-spherical substrate configuration.

The adhesion of a thin sheet to a substrate with a mismatched Gaussian curvature comes with an unavoidable geometry-dependent stress in the sheet \cite{Davidovitch2019}. For example, deforming a flat sheet of radius $W_0$ into a sphere of radius $R_0$ results in a strain $\epsilon_{geom}\sim (W_0/R_0)^2$ and a corresponding in-plane stress $\sigma_{geom}\sim Y(W_0/R_0)^2$ \cite{Bico2018}, where $Y=Eh$ is the sheet's stretching modulus.
An in-plane stress $\sigma_{geom}$ causes a normal force per unit area $P=\sigma_{geom}\kappa_{mean}$, where $\kappa_{mean}$ is the mean curvature of the interface. As the force of friction is proportional to the normal force, we get $m\propto P = \sigma_{geom}\kappa_{mean}$ \cite{elasticity}. Both $\sigma_{geom}$ and $\kappa_{mean}$ have non-zero values in the case of a sphere and hence a non-zero value of $P$. On the other hand, 
$\sigma_{geom}=0$ for a cylinder, and both $\sigma_{geom}=0$ and $\kappa_{mean}=0$ for a flat substrate, leading to $P=0$ in both these cases. This model explains the observation in Fig. \ref{fig:Figure2} and suggests a mechanism through which geometrical incompatibility can affect the force of friction at a soft interface.

\begin{figure}
\includegraphics[width=0.9\linewidth]{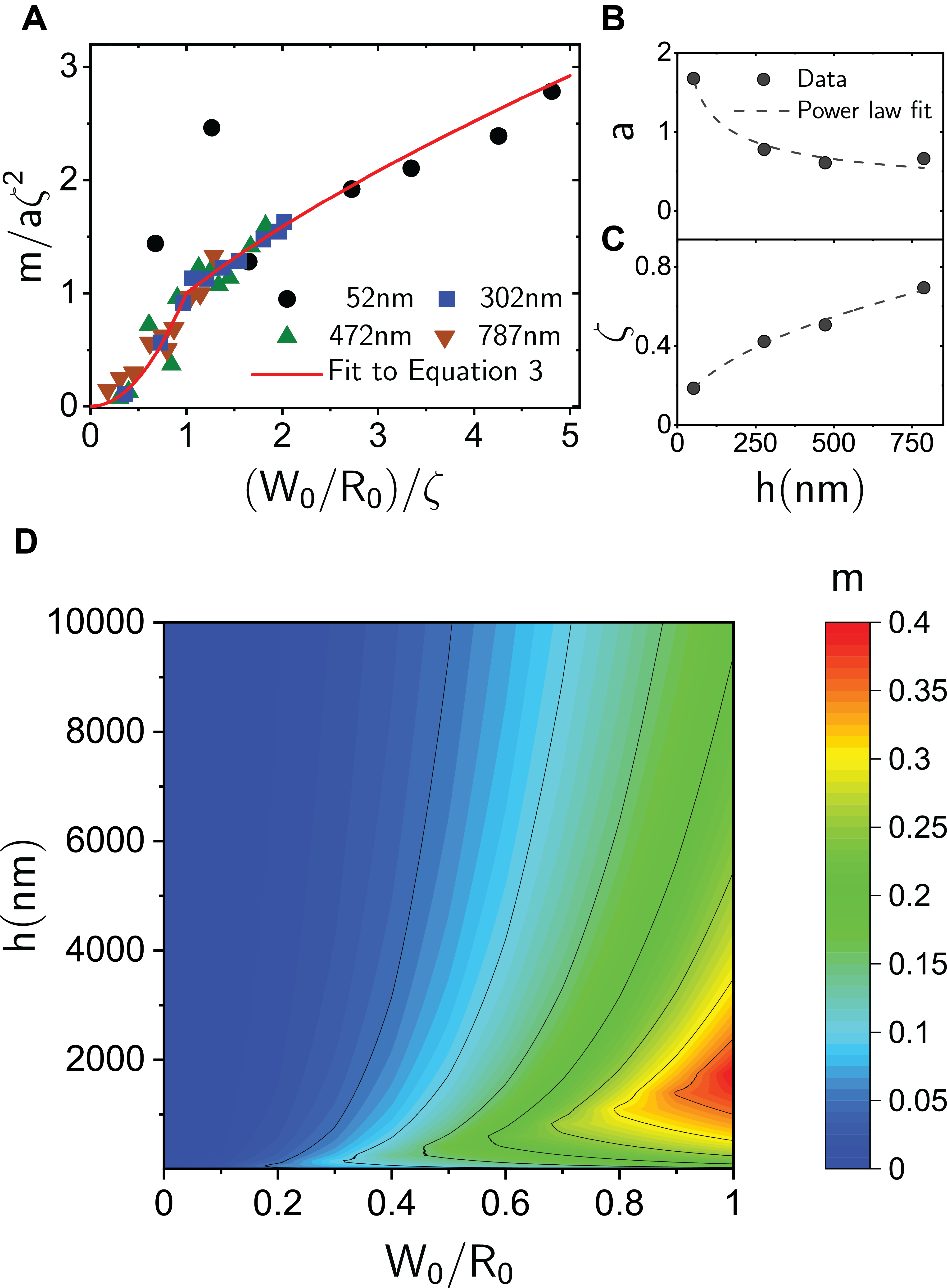} 
\caption{\textbf{Friction phase diagram:} (A) Variation of $m/{a\zeta^2}$ with $\frac{W_0/R_0}{\zeta}$ for flat sheets ($h=52nm$, $302nm$, $472nm$ and  $787nm$) on spherical substrates. All the data collapse onto a single curve when plotted in terms of scaled variables. The red curve is a plot of equation (\ref{eq:slopenorm}). (B) and (C) The variation of the parameters $a$ and $\zeta$, respectively, with $h$, obtained when equation (\ref{eq:slope}) is fitted to the $m(W_0/R_0)$ data for each individual sheet thickness. The dashed curves are fits to the power-laws $y=Ax^\alpha$. (D) Friction phase diagram showing $m$ (color coded) as a function of sheet size $W_0/R_0$ and sheet thickness $h$. The values of $m$ are obtained from equation (\ref{eq:slope}) using $a(h)$ and $\zeta(h)$ as obtained above by fitting.}
\label{fig:Figure3}
\end{figure}

We further establish the validity of our model 
by applying it to understand the dependence of $m$ on $W_0/R_0$ for the flat sheet-spherical substrate system (Fig. \ref{fig:Figure2}A). The in-plane stress 
induced in a flat sheet when it adheres to a spherical substrate has been calculated by solving the F\"{o}ppl-von K\'{a}rm\'{a}n equations, for example in ref. \cite{Hohlfeld2015}. It is found that for small sheets with $W_0/R_0<\zeta$, where $\zeta$ represents the threshold value of $W_0/R_0$, the stress is tensile everywhere in the sheet. However, for sheets with $W_0/R_0>\zeta$, while the stress remains tensile near the center ($0<r<L$), a compressive zone develops near the edge of the sheet ($L<r<W$). Further, the compressive stress can relax by out-of-plane buckling, e.g., through the formation of wrinkles, etc., which costs negligible energy for very thin sheets. As shown in ref. \cite{Hohlfeld2015} the length $L$ scales with the sheet size as $\frac{L}{R}\sim (\frac{W}{R})^{1/3}$. We can, therefore, write the following expression for 
strain averaged over 
sheet:

\begin{equation}
\bar{\epsilon}_{geom}\sim \begin{cases}
          (\frac{W_0}{R_0})^2 &   \frac{W_0}{R_0}\leq \zeta \\
          (\frac{L}{R_0})^2 \sim (\frac{W_0}{R_0})^{2/3} &   \frac{W_0}{R_0}>\zeta \\
     \end{cases}
     \label{eq:strain}
\end{equation}

Corresponding to the strain $\bar{\epsilon}_{geom}$, there would be a stress $\bar{\sigma}_{geom}=Y \bar{\epsilon}_{geom}$ and a corresponding 
pressure $P=\frac{\bar{\sigma}_{geom}}{R}$. Further, since $m\propto P$, we expect the following functional form for the variation of $m$ with the sheet size for the same value of mean curvature:

 \begin{equation}
m=\begin{cases}
         a (\frac{W_0}{R_0})^2    & \:\frac{W_0}{R_0}\leq\zeta\\
          a \zeta^{4/3}(\frac{W_0}{R_0})^{2/3}     &\:     \frac{W_0}{R_0}>\zeta \\
     \end{cases}
     \label{eq:slope}
\end{equation}

Here, $a$ is a constant that may depend on the elastic, frictional and geometrical properties of the interface. The coefficient of the term corresponding to $\frac{W_0}{R_0}>\zeta$ has been obtained by assuming that $m$ is continuous across $\frac{W_0}{R_0}=\zeta$. 
According to equation (\ref{eq:slope}) $m$ has a cusp at $W_0/R_0=\zeta$, signifying that the 
frictional behavior 
undergoes a transition at the sheet radius above which the compressive zone appears near the edge of the sheet. We fit equation (\ref{eq:slope}) to our $m(W_0/R_0)$ data with $\zeta$ and $a$ as free parameters. 
The function fits the data well (red line in Fig. \ref{fig:Figure2}A).

We repeat the experiment with sheets of thicknesses: $h=52nm$, $302nm$, $472nm$ and  $787nm$ and fit equation (\ref{eq:slope}) to the $m(W_0/R_0)$ data for each value of $h$. Values of the fitting parameters $a$ and $\zeta$, so obtained, are plotted as functions of $h$ in Fig. \ref{fig:Figure3}B,C. Since $m(W_0/R_0)$ has a cusp at $W_0/R_0=\zeta$, where $m=a\zeta^2$, we scale $W_0/R_0$ with $\zeta$ and $m$ with $a\zeta^2$ and plot the data for various thicknesses together in Fig. \ref{fig:Figure3}A. The data for different thicknesses collapse onto a single curve. The scaled version of equation (\ref{eq:slope}): 
\begin{equation}
    \frac{m}{a\zeta^2}=\begin{cases}
         (\frac{W_0/R_0}{\zeta})^2    & \:\frac{W_0/R_0}{\zeta}\leq 1\\
          (\frac{W_0/R_0}{\zeta})^{2/3}     &\:     \frac{W_0/R_0}{\zeta}>1 \\
     \end{cases}
     \label{eq:slopenorm}
\end{equation} 
is plotted as a red curve in Fig. \ref{fig:Figure3}A and fits the data well.

Both $a$ and $\zeta$ show dependence on 
$h$ (Fig. \ref{fig:Figure3}B,C), which we model by power-law functions $y=Ax^\alpha$ and find by fitting that $\zeta \sim 0.027 h^{0.48}$ and $a \sim 8.3h^{-0.41}$, shown as dashed curves in the respective graphs. Since $m$ depends both on $h$
, and $W_0/R_0$, 
in order to obtain a comprehensive picture 
we make a 2D color plot, as shown in Fig. \ref{fig:Figure3}D, by computing $m$ using equation (\ref{eq:slope}) and the power-law fits for $a(h)$ and $\zeta(h)$ obtained above. We find, as also noted before, that $m$ increases monotonically with $W_0/R_0$. On the other hand, $m$ shows a non-monotonic dependence on sheet thickness: it is small for both very small values of $h$ as well as for large values of $h$, reaching a maximum for an intermediate value, e.g., for $W_0/R_0\sim1$, $m$ is maximum around $h\sim 1.5 \mu m$. The small value of $m$ for very thin sheets is due to their low stretching modulus $Y$, which results in smaller $\sigma_{geom}$. 
On the other hand, for very thick sheets, the surface energy may be insufficient to cause the elastic deformation required to conform the sheet to the incompatible geometry of the substrate. This observation highlights that although the effect of the frictional mechanism studied here may be negligible for rigid systems, it may play an important role in a large class of soft systems.


Friction is ubiquitous in any dynamical phenomena involving relative motion at an interface and plays a particularly important role in soft systems, where it shows certain unique features not seen typically in rigid systems \cite{hsia2021rougher}. This paper presents a simple experimental technique to study friction at a thin sheet-hydrogel interface. Our study reveals a strong dependence of friction on geometry. In particular, friction is significantly modified when the thin sheet and the substrate have incompatible geometries, viz., a mismatch in Gaussian curvatures. A thin sheet under geometrically incompatible confinement has become a prototypical system to study the emergence of rich patterns of wrinkles, crumples, and folds due to the unavoidable and often non-trivial stress distribution caused by confinement. The present work demonstrates for the first time that such a stress distribution can have a significant effect on the frictional response of the system as well. The insight obtained from the present study can have important implications for our understanding of friction in many biological, nano-scale, or other soft systems. For example, it motivates us to re-examine our understanding of phenomena such as the curvature dependence of the migration of cells called ‘curvotaxis’ \cite{Pieuchot2018} and the dependence of friction on roughness at soft interfaces \cite{hsia2021rougher}.

\section*{Methods}

\subsection*{Preparation of thin sheets}
Thin sheets are prepared by spin coating a dilute solution of polystyrene (average $M_w=192K$, Sigma-Aldrich) and a small amount ($\sim 0.002 \%$) of fluorescent dye Nile-red (Sigma-Aldrich, $72485$) in toluene (anhydrous $99.8\%$, Sigma-Aldrich, $244511$). The solution is thoroughly mixed and filtered before spin coating on clean microscope glass slides. Sheets of different thicknesses are obtained by varying the concentration of polystyrene between $1 wt\%$ to $9 wt\%$. Circular discs of radius $W_0=0.75mm$ to $5mm$ are cut near the center of the slide, where the thickness is uniform, and first floated on the surface of a deionized water bath and later transferred to the swollen hydrogel substrate.

\subsection*{Preparation of hydrogel substrates}
Commercially available polyacrylamide particles (diameter $\sim 2mm$ in the dry state) are soaked in deionized water to obtain the swollen hydrogel spheres of diameter $\sim 12 mm$, which are used as the spherical substrates. 
The swollen hydrogel spheres have Young's modulus of $114 KPa$, as measured using UTM, and an extremely smooth surface (optical micrograph in SI). Flat substrates are obtained by cutting the swollen hydrogel spheres using a sharp, flat blade. Cylindrical substrates are obtained by cutting the swollen hydrogel spheres with a sharp blade bent into a cylindrical shape. Information regarding the characterization of the geometry and roughness of the cylindrical surfaces so obtained is provided in the SI. 
We notice that the surfaces obtained by cutting are rougher than the surfaces of the original hydrogel spheres. 

\subsection*{Imaging setup}
The imaging setup is shown schematically in Fig. \ref{fig:Figure1}A.  We image the sheet and the substrate separately using two different imaging schemes. While a brightfield illumination in the transmission mode is used to image the hydrogel substrate, a fluorescence-based technique is used to image the sheet. 
A green laser ($\lambda=532nm$) is used to excite fluorescence in the Nile red dye present in the polystyrene sheet, which is captured using the camera after filtering out the green light using a low pass filter. The two different imaging schemes are implemented in sequence, one after the other. 
Pairs of such images are recorded at equal intervals of time ($\Delta t \sim 10 \:minutes$).

\subsection*{Determination of the sheet and substrate dimensions}
We determine the radii of the sheet $W(t)$ and the hydrogel substrate $R(t)$ from the fluorescence and bright field images, respectively, by using image analysis codes written in Python. 

The size of the substrate at any instant of time is determined from the bright-field image, for example, as shown in Fig. \ref{fig:Figure2}B. The red channel of the image is thresholded, and a connected component analysis is run to identify the substrate in the image. While the dimension of the best-fit ellipse is used in the case of a spherical and a flat substrate, the dimension of the best-fit bounding rectangle is used in the case of a cylindrical substrate. Although the eccentricity observed in the case of a flat/spherical substrate is usually very small, the quality of fit is better for an ellipse than for a circle.  

The size of the sheet 
is determined from the red channel of the fluorescent image in a two-step process. In the first step, a rough estimate of the sheet center is made by thresholding the image, running a connected component analysis, and finding the best-fit ellipse. In the second stage, we obtain a refined measure of the sheet center and radius by analysing its radial profile. 
Using the center obtained in the first step as the origin 
we calculate the radial intensity profile $\langle I\rangle(r)=\frac{1}{2\pi}\int_{0}^{2\pi}I(r,\theta)d\theta$, where $r$ is the distance from the center and $\theta$ is the polar angle. We then convolve $\langle I\rangle (r)$ with the derivative of a Gaussian to find a smooth derivative $I'(r)$ of the function$\langle I\rangle (r)$. In the convolution, we note the value ($I'_{max}$) and position of the peak ($R$). The process is now repeated by taking different center positions in a small neighborhood of the center position estimated in the first step, and we look at the variation in $I'_{max}$ with the center position. We use the center and the radius corresponding to the peak value of $I'_{max}$ to determine the correct center and radius of the sheet with sub-pixel resolution. While the above method is used for spherical and flat substrates, for cylindrical substrates where the response along the curved and axial directions may be different, we measure lengths along the two directions by fitting an ellipse to the sheet image using the connected component analysis.

\section*{Acknowledgements}
We thank Swadhin Agarwal for some early experiments. This research was supported in part by the IIT Delhi New Faculty Seed Grant (DK), SERB, India under the grant SRG/2019/000949 (DK). AC acknowledges UGG, India for research fellowship.

\section*{References}

\bibliography{reference}

\begin{thebibliography}{25}%
\makeatletter
\providecommand \@ifxundefined [1]{%
 \@ifx{#1\undefined}
}%
\providecommand \@ifnum [1]{%
 \ifnum #1\expandafter \@firstoftwo
 \else \expandafter \@secondoftwo
 \fi
}%
\providecommand \@ifx [1]{%
 \ifx #1\expandafter \@firstoftwo
 \else \expandafter \@secondoftwo
 \fi
}%
\providecommand \natexlab [1]{#1}%
\providecommand \enquote  [1]{``#1''}%
\providecommand \bibnamefont  [1]{#1}%
\providecommand \bibfnamefont [1]{#1}%
\providecommand \citenamefont [1]{#1}%
\providecommand \href@noop [0]{\@secondoftwo}%
\providecommand \href [0]{\begingroup \@sanitize@url \@href}%
\providecommand \@href[1]{\@@startlink{#1}\@@href}%
\providecommand \@@href[1]{\endgroup#1\@@endlink}%
\providecommand \@sanitize@url [0]{\catcode `\\12\catcode `\$12\catcode `\&12\catcode `\#12\catcode `\^12\catcode `\_12\catcode `\%12\relax}%
\providecommand \@@startlink[1]{}%
\providecommand \@@endlink[0]{}%
\providecommand \url  [0]{\begingroup\@sanitize@url \@url }%
\providecommand \@url [1]{\endgroup\@href {#1}{\urlprefix }}%
\providecommand \urlprefix  [0]{URL }%
\providecommand \Eprint [0]{\href }%
\providecommand \doibase [0]{https://doi.org/}%
\providecommand \selectlanguage [0]{\@gobble}%
\providecommand \bibinfo  [0]{\@secondoftwo}%
\providecommand \bibfield  [0]{\@secondoftwo}%
\providecommand \translation [1]{[#1]}%
\providecommand \BibitemOpen [0]{}%
\providecommand \bibitemStop [0]{}%
\providecommand \bibitemNoStop [0]{.\EOS\space}%
\providecommand \EOS [0]{\spacefactor3000\relax}%
\providecommand \BibitemShut  [1]{\csname bibitem#1\endcsname}%
\let\auto@bib@innerbib\@empty
\bibitem [{\citenamefont {Hure}\ \emph {et~al.}(2012)\citenamefont {Hure}, \citenamefont {Roman},\ and\ \citenamefont {Bico}}]{Hure2012}%
  \BibitemOpen
  \bibfield  {author} {\bibinfo {author} {\bibfnamefont {J.}~\bibnamefont {Hure}}, \bibinfo {author} {\bibfnamefont {B.}~\bibnamefont {Roman}},\ and\ \bibinfo {author} {\bibfnamefont {J.}~\bibnamefont {Bico}},\ }\bibfield  {title} {\bibinfo {title} {Stamping and wrinkling of elastic plates},\ }\href@noop {} {\bibfield  {journal} {\bibinfo  {journal} {Phys. Rev. Lett.}\ }\textbf {\bibinfo {volume} {109}},\ \bibinfo {pages} {054302} (\bibinfo {year} {2012})}\BibitemShut {NoStop}%
\bibitem [{\citenamefont {Hure}\ \emph {et~al.}(2011)\citenamefont {Hure}, \citenamefont {Roman},\ and\ \citenamefont {Bico}}]{Hure2011}%
  \BibitemOpen
  \bibfield  {author} {\bibinfo {author} {\bibfnamefont {J.}~\bibnamefont {Hure}}, \bibinfo {author} {\bibfnamefont {B.}~\bibnamefont {Roman}},\ and\ \bibinfo {author} {\bibfnamefont {J.}~\bibnamefont {Bico}},\ }\bibfield  {title} {\bibinfo {title} {Wrapping an adhesive sphere with an elastic sheet},\ }\href@noop {} {\bibfield  {journal} {\bibinfo  {journal} {Phys. Rev. Lett.}\ }\textbf {\bibinfo {volume} {106}},\ \bibinfo {pages} {174301} (\bibinfo {year} {2011})}\BibitemShut {NoStop}%
\bibitem [{\citenamefont {King}\ \emph {et~al.}(2012)\citenamefont {King}, \citenamefont {Schroll}, \citenamefont {Davidovitch},\ and\ \citenamefont {Menon}}]{King2012}%
  \BibitemOpen
  \bibfield  {author} {\bibinfo {author} {\bibfnamefont {H.}~\bibnamefont {King}}, \bibinfo {author} {\bibfnamefont {R.~D.}\ \bibnamefont {Schroll}}, \bibinfo {author} {\bibfnamefont {B.}~\bibnamefont {Davidovitch}},\ and\ \bibinfo {author} {\bibfnamefont {N.}~\bibnamefont {Menon}},\ }\bibfield  {title} {\bibinfo {title} {Elastic sheet on a liquid drop reveals wrinkling and crumpling as distinct symmetry-breaking instabilities},\ }\href@noop {} {\bibfield  {journal} {\bibinfo  {journal} {Proc. Natl. Acad. Sci. U.S.A.}\ }\textbf {\bibinfo {volume} {109}},\ \bibinfo {pages} {9716} (\bibinfo {year} {2012})}\BibitemShut {NoStop}%
\bibitem [{\citenamefont {Tobasco}\ \emph {et~al.}(2022)\citenamefont {Tobasco}, \citenamefont {Timounay}, \citenamefont {Todorova}, \citenamefont {Leggat}, \citenamefont {Paulsen},\ and\ \citenamefont {Katifori}}]{tobasco2022exact}%
  \BibitemOpen
  \bibfield  {author} {\bibinfo {author} {\bibfnamefont {I.}~\bibnamefont {Tobasco}}, \bibinfo {author} {\bibfnamefont {Y.}~\bibnamefont {Timounay}}, \bibinfo {author} {\bibfnamefont {D.}~\bibnamefont {Todorova}}, \bibinfo {author} {\bibfnamefont {G.~C.}\ \bibnamefont {Leggat}}, \bibinfo {author} {\bibfnamefont {J.~D.}\ \bibnamefont {Paulsen}},\ and\ \bibinfo {author} {\bibfnamefont {E.}~\bibnamefont {Katifori}},\ }\bibfield  {title} {\bibinfo {title} {Exact solutions for the wrinkle patterns of confined elastic shells},\ }\href@noop {} {\bibfield  {journal} {\bibinfo  {journal} {Nature Physics}\ }\textbf {\bibinfo {volume} {18}},\ \bibinfo {pages} {1099} (\bibinfo {year} {2022})}\BibitemShut {NoStop}%
\bibitem [{\citenamefont {Gong}\ \emph {et~al.}(1997)\citenamefont {Gong}, \citenamefont {Higa}, \citenamefont {Iwasaki}, \citenamefont {Katsuyama},\ and\ \citenamefont {Osada}}]{gong1997friction}%
  \BibitemOpen
  \bibfield  {author} {\bibinfo {author} {\bibfnamefont {J.}~\bibnamefont {Gong}}, \bibinfo {author} {\bibfnamefont {M.}~\bibnamefont {Higa}}, \bibinfo {author} {\bibfnamefont {Y.}~\bibnamefont {Iwasaki}}, \bibinfo {author} {\bibfnamefont {Y.}~\bibnamefont {Katsuyama}},\ and\ \bibinfo {author} {\bibfnamefont {Y.}~\bibnamefont {Osada}},\ }\bibfield  {title} {\bibinfo {title} {Friction of gels},\ }\href@noop {} {\bibfield  {journal} {\bibinfo  {journal} {The Journal of Physical Chemistry B}\ }\textbf {\bibinfo {volume} {101}},\ \bibinfo {pages} {5487} (\bibinfo {year} {1997})}\BibitemShut {NoStop}%
\bibitem [{\citenamefont {Cuccia}\ \emph {et~al.}(2020)\citenamefont {Cuccia}, \citenamefont {Pothineni}, \citenamefont {Wu}, \citenamefont {M{\'e}ndez~Harper},\ and\ \citenamefont {Burton}}]{Cuccia2020}%
  \BibitemOpen
  \bibfield  {author} {\bibinfo {author} {\bibfnamefont {N.~L.}\ \bibnamefont {Cuccia}}, \bibinfo {author} {\bibfnamefont {S.}~\bibnamefont {Pothineni}}, \bibinfo {author} {\bibfnamefont {B.}~\bibnamefont {Wu}}, \bibinfo {author} {\bibfnamefont {J.}~\bibnamefont {M{\'e}ndez~Harper}},\ and\ \bibinfo {author} {\bibfnamefont {J.~C.}\ \bibnamefont {Burton}},\ }\bibfield  {title} {\bibinfo {title} {Pore-size dependence and slow relaxation of hydrogel friction on smooth surfaces},\ }\href@noop {} {\bibfield  {journal} {\bibinfo  {journal} {Proc. Natl. Acad. Sci. U.S.A.}\ }\textbf {\bibinfo {volume} {117}},\ \bibinfo {pages} {11247} (\bibinfo {year} {2020})}\BibitemShut {NoStop}%
\bibitem [{\citenamefont {Hohlfeld}\ and\ \citenamefont {Davidovitch}(2015)}]{Hohlfeld2015}%
  \BibitemOpen
  \bibfield  {author} {\bibinfo {author} {\bibfnamefont {E.}~\bibnamefont {Hohlfeld}}\ and\ \bibinfo {author} {\bibfnamefont {B.}~\bibnamefont {Davidovitch}},\ }\bibfield  {title} {\bibinfo {title} {Sheet on a deformable sphere: Wrinkle patterns suppress curvature-induced delamination},\ }\href@noop {} {\bibfield  {journal} {\bibinfo  {journal} {Phys. Rev. E}\ }\textbf {\bibinfo {volume} {91}},\ \bibinfo {pages} {012407} (\bibinfo {year} {2015})}\BibitemShut {NoStop}%
\bibitem [{\citenamefont {Pieuchot}\ \emph {et~al.}(2018)\citenamefont {Pieuchot}, \citenamefont {Marteau}, \citenamefont {Guignandon}, \citenamefont {Dos~Santos}, \citenamefont {Brigaud}, \citenamefont {Chauvy}, \citenamefont {Cloatre}, \citenamefont {Ponche}, \citenamefont {Petithory}, \citenamefont {Rougerie}, \citenamefont {Vassaux}, \citenamefont {Milan}, \citenamefont {Tusamda~Wakhloo}, \citenamefont {Spangenberg}, \citenamefont {Bigerelle},\ and\ \citenamefont {Anselme}}]{Pieuchot2018}%
  \BibitemOpen
  \bibfield  {author} {\bibinfo {author} {\bibfnamefont {L.}~\bibnamefont {Pieuchot}}, \bibinfo {author} {\bibfnamefont {J.}~\bibnamefont {Marteau}}, \bibinfo {author} {\bibfnamefont {A.}~\bibnamefont {Guignandon}}, \bibinfo {author} {\bibfnamefont {T.}~\bibnamefont {Dos~Santos}}, \bibinfo {author} {\bibfnamefont {I.}~\bibnamefont {Brigaud}}, \bibinfo {author} {\bibfnamefont {P.-F.}\ \bibnamefont {Chauvy}}, \bibinfo {author} {\bibfnamefont {T.}~\bibnamefont {Cloatre}}, \bibinfo {author} {\bibfnamefont {A.}~\bibnamefont {Ponche}}, \bibinfo {author} {\bibfnamefont {T.}~\bibnamefont {Petithory}}, \bibinfo {author} {\bibfnamefont {P.}~\bibnamefont {Rougerie}}, \bibinfo {author} {\bibfnamefont {M.}~\bibnamefont {Vassaux}}, \bibinfo {author} {\bibfnamefont {J.-L.}\ \bibnamefont {Milan}}, \bibinfo {author} {\bibfnamefont {N.}~\bibnamefont {Tusamda~Wakhloo}}, \bibinfo {author} {\bibfnamefont {A.}~\bibnamefont {Spangenberg}}, \bibinfo {author} {\bibfnamefont {M.}~\bibnamefont {Bigerelle}},\ and\ \bibinfo {author}
  {\bibfnamefont {K.}~\bibnamefont {Anselme}},\ }\bibfield  {title} {\bibinfo {title} {Curvotaxis directs cell migration through cell-scale curvature landscapes},\ }\href@noop {} {\bibfield  {journal} {\bibinfo  {journal} {Nature Communications}\ }\textbf {\bibinfo {volume} {9}},\ \bibinfo {pages} {3995} (\bibinfo {year} {2018})}\BibitemShut {NoStop}%
\bibitem [{\citenamefont {Takezoe}\ and\ \citenamefont {Takanishi}(2006)}]{2006JaJAP..45..597T}%
  \BibitemOpen
  \bibfield  {author} {\bibinfo {author} {\bibfnamefont {H.}~\bibnamefont {Takezoe}}\ and\ \bibinfo {author} {\bibfnamefont {Y.}~\bibnamefont {Takanishi}},\ }\bibfield  {title} {\bibinfo {title} {Bent-core liquid crystals: their mysterious and attractive world},\ }\href@noop {} {\bibfield  {journal} {\bibinfo  {journal} {Japanese journal of applied physics}\ }\textbf {\bibinfo {volume} {45}},\ \bibinfo {pages} {597} (\bibinfo {year} {2006})}\BibitemShut {NoStop}%
\bibitem [{\citenamefont {Niv}\ and\ \citenamefont {Efrati}(2018)}]{Niv2018}%
  \BibitemOpen
  \bibfield  {author} {\bibinfo {author} {\bibfnamefont {I.}~\bibnamefont {Niv}}\ and\ \bibinfo {author} {\bibfnamefont {E.}~\bibnamefont {Efrati}},\ }\bibfield  {title} {\bibinfo {title} {Geometric frustration and compatibility conditions for two-dimensional director fields},\ }\href@noop {} {\bibfield  {journal} {\bibinfo  {journal} {Soft matter}\ }\textbf {\bibinfo {volume} {14}},\ \bibinfo {pages} {424} (\bibinfo {year} {2018})}\BibitemShut {NoStop}%
\bibitem [{\citenamefont {Lenz}\ and\ \citenamefont {Witten}(2017)}]{Lenz2017}%
  \BibitemOpen
  \bibfield  {author} {\bibinfo {author} {\bibfnamefont {M.}~\bibnamefont {Lenz}}\ and\ \bibinfo {author} {\bibfnamefont {T.~A.}\ \bibnamefont {Witten}},\ }\bibfield  {title} {\bibinfo {title} {Geometrical frustration yields fibre formation in self-assembly},\ }\href@noop {} {\bibfield  {journal} {\bibinfo  {journal} {Nature physics}\ }\textbf {\bibinfo {volume} {13}},\ \bibinfo {pages} {1100} (\bibinfo {year} {2017})}\BibitemShut {NoStop}%
\bibitem [{\citenamefont {Struik}(1988)}]{Struik1988}%
  \BibitemOpen
  \bibfield  {author} {\bibinfo {author} {\bibfnamefont {D.~J.}\ \bibnamefont {Struik}},\ }\href@noop {} {\emph {\bibinfo {title} {Lectures on classical differential geometry}}}\ (\bibinfo  {publisher} {Dover Publications},\ \bibinfo {year} {1988})\ p.\ \bibinfo {pages} {232}\BibitemShut {NoStop}%
\bibitem [{\citenamefont {Urbakh}\ \emph {et~al.}(2004)\citenamefont {Urbakh}, \citenamefont {Klafter}, \citenamefont {Gourdon},\ and\ \citenamefont {Israelachvili}}]{urbakh2004nonlinear}%
  \BibitemOpen
  \bibfield  {author} {\bibinfo {author} {\bibfnamefont {M.}~\bibnamefont {Urbakh}}, \bibinfo {author} {\bibfnamefont {J.}~\bibnamefont {Klafter}}, \bibinfo {author} {\bibfnamefont {D.}~\bibnamefont {Gourdon}},\ and\ \bibinfo {author} {\bibfnamefont {J.}~\bibnamefont {Israelachvili}},\ }\bibfield  {title} {\bibinfo {title} {The nonlinear nature of friction},\ }\href@noop {} {\bibfield  {journal} {\bibinfo  {journal} {Nature}\ }\textbf {\bibinfo {volume} {430}},\ \bibinfo {pages} {525} (\bibinfo {year} {2004})}\BibitemShut {NoStop}%
\bibitem [{\citenamefont {Vanossi}\ \emph {et~al.}(2013)\citenamefont {Vanossi}, \citenamefont {Manini}, \citenamefont {Urbakh}, \citenamefont {Zapperi},\ and\ \citenamefont {Tosatti}}]{vanossi2013colloquium}%
  \BibitemOpen
  \bibfield  {author} {\bibinfo {author} {\bibfnamefont {A.}~\bibnamefont {Vanossi}}, \bibinfo {author} {\bibfnamefont {N.}~\bibnamefont {Manini}}, \bibinfo {author} {\bibfnamefont {M.}~\bibnamefont {Urbakh}}, \bibinfo {author} {\bibfnamefont {S.}~\bibnamefont {Zapperi}},\ and\ \bibinfo {author} {\bibfnamefont {E.}~\bibnamefont {Tosatti}},\ }\bibfield  {title} {\bibinfo {title} {Colloquium: Modeling friction: From nanoscale to mesoscale},\ }\href@noop {} {\bibfield  {journal} {\bibinfo  {journal} {Reviews of Modern Physics}\ }\textbf {\bibinfo {volume} {85}},\ \bibinfo {pages} {529} (\bibinfo {year} {2013})}\BibitemShut {NoStop}%
\bibitem [{\citenamefont {Vanossi}\ \emph {et~al.}(2020)\citenamefont {Vanossi}, \citenamefont {Bechinger},\ and\ \citenamefont {Urbakh}}]{Vanossi2020}%
  \BibitemOpen
  \bibfield  {author} {\bibinfo {author} {\bibfnamefont {A.}~\bibnamefont {Vanossi}}, \bibinfo {author} {\bibfnamefont {C.}~\bibnamefont {Bechinger}},\ and\ \bibinfo {author} {\bibfnamefont {M.}~\bibnamefont {Urbakh}},\ }\bibfield  {title} {\bibinfo {title} {{Structural lubricity in soft and hard matter systems}},\ }\href {https://doi.org/10.1038/s41467-020-18429-1} {\bibfield  {journal} {\bibinfo  {journal} {Nat. Commun.}\ }\textbf {\bibinfo {volume} {11}},\ \bibinfo {pages} {4657} (\bibinfo {year} {2020})}\BibitemShut {NoStop}%
\bibitem [{\citenamefont {Martin}\ and\ \citenamefont {Erdemir}(2018)}]{superlubricity}%
  \BibitemOpen
  \bibfield  {author} {\bibinfo {author} {\bibfnamefont {J.~M.}\ \bibnamefont {Martin}}\ and\ \bibinfo {author} {\bibfnamefont {A.}~\bibnamefont {Erdemir}},\ }\bibfield  {title} {\bibinfo {title} {Superlubricity: Friction’s vanishing act},\ }\href@noop {} {\bibfield  {journal} {\bibinfo  {journal} {Physics Today}\ }\textbf {\bibinfo {volume} {71}},\ \bibinfo {pages} {40} (\bibinfo {year} {2018})}\BibitemShut {NoStop}%
\bibitem [{\citenamefont {Simi{\v{c}}}\ \emph {et~al.}(2020)\citenamefont {Simi{\v{c}}}, \citenamefont {Yetkin}, \citenamefont {Zhang},\ and\ \citenamefont {Spencer}}]{Simič2020}%
  \BibitemOpen
  \bibfield  {author} {\bibinfo {author} {\bibfnamefont {R.}~\bibnamefont {Simi{\v{c}}}}, \bibinfo {author} {\bibfnamefont {M.}~\bibnamefont {Yetkin}}, \bibinfo {author} {\bibfnamefont {K.}~\bibnamefont {Zhang}},\ and\ \bibinfo {author} {\bibfnamefont {N.~D.}\ \bibnamefont {Spencer}},\ }\bibfield  {title} {\bibinfo {title} {Importance of hydration and surface structure for friction of acrylamide hydrogels},\ }\href@noop {} {\bibfield  {journal} {\bibinfo  {journal} {Tribology Letters}\ }\textbf {\bibinfo {volume} {68}},\ \bibinfo {pages} {1} (\bibinfo {year} {2020})}\BibitemShut {NoStop}%
\bibitem [{\citenamefont {Wu}\ \emph {et~al.}(2021)\citenamefont {Wu}, \citenamefont {Harper},\ and\ \citenamefont {Burton}}]{wu2021relaxation}%
  \BibitemOpen
  \bibfield  {author} {\bibinfo {author} {\bibfnamefont {B.}~\bibnamefont {Wu}}, \bibinfo {author} {\bibfnamefont {J.~M.}\ \bibnamefont {Harper}},\ and\ \bibinfo {author} {\bibfnamefont {J.~C.}\ \bibnamefont {Burton}},\ }\bibfield  {title} {\bibinfo {title} {Relaxation and recovery in hydrogel friction on smooth surfaces},\ }\href@noop {} {\bibfield  {journal} {\bibinfo  {journal} {Experimental Mechanics}\ }\textbf {\bibinfo {volume} {61}},\ \bibinfo {pages} {1081} (\bibinfo {year} {2021})}\BibitemShut {NoStop}%
\bibitem [{\citenamefont {Matan}\ \emph {et~al.}(2002)\citenamefont {Matan}, \citenamefont {Williams}, \citenamefont {Witten},\ and\ \citenamefont {Nagel}}]{Matan2002}%
  \BibitemOpen
  \bibfield  {author} {\bibinfo {author} {\bibfnamefont {K.}~\bibnamefont {Matan}}, \bibinfo {author} {\bibfnamefont {R.~B.}\ \bibnamefont {Williams}}, \bibinfo {author} {\bibfnamefont {T.~A.}\ \bibnamefont {Witten}},\ and\ \bibinfo {author} {\bibfnamefont {S.~R.}\ \bibnamefont {Nagel}},\ }\bibfield  {title} {\bibinfo {title} {Crumpling a thin sheet},\ }\href@noop {} {\bibfield  {journal} {\bibinfo  {journal} {Phys. Rev. Lett.}\ }\textbf {\bibinfo {volume} {88}},\ \bibinfo {pages} {076101} (\bibinfo {year} {2002})}\BibitemShut {NoStop}%
\bibitem [{\citenamefont {Lomholt}\ \emph {et~al.}(2013)\citenamefont {Lomholt}, \citenamefont {Lizana}, \citenamefont {Metzler},\ and\ \citenamefont {Ambj\"ornsson}}]{Lomholt2013}%
  \BibitemOpen
  \bibfield  {author} {\bibinfo {author} {\bibfnamefont {M.~A.}\ \bibnamefont {Lomholt}}, \bibinfo {author} {\bibfnamefont {L.}~\bibnamefont {Lizana}}, \bibinfo {author} {\bibfnamefont {R.}~\bibnamefont {Metzler}},\ and\ \bibinfo {author} {\bibfnamefont {T.}~\bibnamefont {Ambj\"ornsson}},\ }\bibfield  {title} {\bibinfo {title} {Microscopic origin of the logarithmic time evolution of aging processes in complex systems},\ }\href@noop {} {\bibfield  {journal} {\bibinfo  {journal} {Phys. Rev. Lett.}\ }\textbf {\bibinfo {volume} {110}},\ \bibinfo {pages} {208301} (\bibinfo {year} {2013})}\BibitemShut {NoStop}%
\bibitem [{\citenamefont {Gong}(2006)}]{gong2006friction}%
  \BibitemOpen
  \bibfield  {author} {\bibinfo {author} {\bibfnamefont {J.~P.}\ \bibnamefont {Gong}},\ }\bibfield  {title} {\bibinfo {title} {Friction and lubrication of hydrogels—its richness and complexity},\ }\href@noop {} {\bibfield  {journal} {\bibinfo  {journal} {Soft matter}\ }\textbf {\bibinfo {volume} {2}},\ \bibinfo {pages} {544} (\bibinfo {year} {2006})}\BibitemShut {NoStop}%
\bibitem [{\citenamefont {Davidovitch}\ \emph {et~al.}(2019)\citenamefont {Davidovitch}, \citenamefont {Sun},\ and\ \citenamefont {Grason}}]{Davidovitch2019}%
  \BibitemOpen
  \bibfield  {author} {\bibinfo {author} {\bibfnamefont {B.}~\bibnamefont {Davidovitch}}, \bibinfo {author} {\bibfnamefont {Y.}~\bibnamefont {Sun}},\ and\ \bibinfo {author} {\bibfnamefont {G.~M.}\ \bibnamefont {Grason}},\ }\bibfield  {title} {\bibinfo {title} {Geometrically incompatible confinement of solids},\ }\href@noop {} {\bibfield  {journal} {\bibinfo  {journal} {Proc. Natl. Acad. Sci. U.S.A.}\ }\textbf {\bibinfo {volume} {116}},\ \bibinfo {pages} {1483} (\bibinfo {year} {2019})}\BibitemShut {NoStop}%
\bibitem [{\citenamefont {Bico}\ \emph {et~al.}(2018)\citenamefont {Bico}, \citenamefont {Reyssat},\ and\ \citenamefont {Roman}}]{Bico2018}%
  \BibitemOpen
  \bibfield  {author} {\bibinfo {author} {\bibfnamefont {J.}~\bibnamefont {Bico}}, \bibinfo {author} {\bibfnamefont {{\'E}.}~\bibnamefont {Reyssat}},\ and\ \bibinfo {author} {\bibfnamefont {B.}~\bibnamefont {Roman}},\ }\bibfield  {title} {\bibinfo {title} {Elastocapillarity: When surface tension deforms elastic solids},\ }\href@noop {} {\bibfield  {journal} {\bibinfo  {journal} {Annual Review of Fluid Mechanics}\ }\textbf {\bibinfo {volume} {50}},\ \bibinfo {pages} {629} (\bibinfo {year} {2018})}\BibitemShut {NoStop}%
\bibitem [{\citenamefont {Pomeau}\ and\ \citenamefont {Audoly}(2010)}]{elasticity}%
  \BibitemOpen
  \bibfield  {author} {\bibinfo {author} {\bibfnamefont {Y.}~\bibnamefont {Pomeau}}\ and\ \bibinfo {author} {\bibfnamefont {B.}~\bibnamefont {Audoly}},\ }\href@noop {} {\emph {\bibinfo {title} {Elasticity and Geometry: From Hair Curls to the Non-linear Response of Shells}}}\ (\bibinfo  {publisher} {Oxford University Press},\ \bibinfo {year} {2010})\BibitemShut {NoStop}%
\bibitem [{\citenamefont {Hsia}\ \emph {et~al.}(2021)\citenamefont {Hsia}, \citenamefont {Franklin}, \citenamefont {Audebert}, \citenamefont {Brouwer}, \citenamefont {Bonn},\ and\ \citenamefont {Weber}}]{hsia2021rougher}%
  \BibitemOpen
  \bibfield  {author} {\bibinfo {author} {\bibfnamefont {F.-C.}\ \bibnamefont {Hsia}}, \bibinfo {author} {\bibfnamefont {S.}~\bibnamefont {Franklin}}, \bibinfo {author} {\bibfnamefont {P.}~\bibnamefont {Audebert}}, \bibinfo {author} {\bibfnamefont {A.~M.}\ \bibnamefont {Brouwer}}, \bibinfo {author} {\bibfnamefont {D.}~\bibnamefont {Bonn}},\ and\ \bibinfo {author} {\bibfnamefont {B.}~\bibnamefont {Weber}},\ }\bibfield  {title} {\bibinfo {title} {Rougher is more slippery: How adhesive friction decreases with increasing surface roughness due to the suppression of capillary adhesion},\ }\href@noop {} {\bibfield  {journal} {\bibinfo  {journal} {Phys. Rev. Res.}\ }\textbf {\bibinfo {volume} {3}},\ \bibinfo {pages} {043204} (\bibinfo {year} {2021})}\BibitemShut {NoStop}%
\end{thebibliography}%


\begin{thebibliography}{9}%
\makeatletter
\providecommand \@ifxundefined [1]{%
 \@ifx{#1\undefined}
}%
\providecommand \@ifnum [1]{%
 \ifnum #1\expandafter \@firstoftwo
 \else \expandafter \@secondoftwo
 \fi
}%
\providecommand \@ifx [1]{%
 \ifx #1\expandafter \@firstoftwo
 \else \expandafter \@secondoftwo
 \fi
}%
\providecommand \natexlab [1]{#1}%
\providecommand \enquote  [1]{``#1''}%
\providecommand \bibnamefont  [1]{#1}%
\providecommand \bibfnamefont [1]{#1}%
\providecommand \citenamefont [1]{#1}%
\providecommand \href@noop [0]{\@secondoftwo}%
\providecommand \href [0]{\begingroup \@sanitize@url \@href}%
\providecommand \@href[1]{\@@startlink{#1}\@@href}%
\providecommand \@@href[1]{\endgroup#1\@@endlink}%
\providecommand \@sanitize@url [0]{\catcode `\\12\catcode `\$12\catcode `\&12\catcode `\#12\catcode `\^12\catcode `\_12\catcode `\%12\relax}%
\providecommand \@@startlink[1]{}%
\providecommand \@@endlink[0]{}%
\providecommand \url  [0]{\begingroup\@sanitize@url \@url }%
\providecommand \@url [1]{\endgroup\@href {#1}{\urlprefix }}%
\providecommand \urlprefix  [0]{URL }%
\providecommand \Eprint [0]{\href }%
\providecommand \doibase [0]{https://doi.org/}%
\providecommand \selectlanguage [0]{\@gobble}%
\providecommand \bibinfo  [0]{\@secondoftwo}%
\providecommand \bibfield  [0]{\@secondoftwo}%
\providecommand \translation [1]{[#1]}%
\providecommand \BibitemOpen [0]{}%
\providecommand \bibitemStop [0]{}%
\providecommand \bibitemNoStop [0]{.\EOS\space}%
\providecommand \EOS [0]{\spacefactor3000\relax}%
\providecommand \BibitemShut  [1]{\csname bibitem#1\endcsname}%
\let\auto@bib@innerbib\@empty
\bibitem [{\citenamefont {Simi{\v{c}}}\ \emph {et~al.}(2020)\citenamefont {Simi{\v{c}}}, \citenamefont {Yetkin}, \citenamefont {Zhang},\ and\ \citenamefont {Spencer}}]{Simi2020}%
  \BibitemOpen
  \bibfield  {author} {\bibinfo {author} {\bibfnamefont {R.}~\bibnamefont {Simi{\v{c}}}}, \bibinfo {author} {\bibfnamefont {M.}~\bibnamefont {Yetkin}}, \bibinfo {author} {\bibfnamefont {K.}~\bibnamefont {Zhang}},\ and\ \bibinfo {author} {\bibfnamefont {N.~D.}\ \bibnamefont {Spencer}},\ }\bibfield  {title} {\bibinfo {title} {Importance of hydration and surface structure for friction of acrylamide hydrogels},\ }\href@noop {} {\bibfield  {journal} {\bibinfo  {journal} {Tribology Letters}\ }\textbf {\bibinfo {volume} {68}},\ \bibinfo {pages} {1} (\bibinfo {year} {2020})}\BibitemShut {NoStop}%
\bibitem [{\citenamefont {Gong}\ \emph {et~al.}(1997)\citenamefont {Gong}, \citenamefont {Higa}, \citenamefont {Iwasaki}, \citenamefont {Katsuyama},\ and\ \citenamefont {Osada}}]{gong1997friction}%
  \BibitemOpen
  \bibfield  {author} {\bibinfo {author} {\bibfnamefont {J.}~\bibnamefont {Gong}}, \bibinfo {author} {\bibfnamefont {M.}~\bibnamefont {Higa}}, \bibinfo {author} {\bibfnamefont {Y.}~\bibnamefont {Iwasaki}}, \bibinfo {author} {\bibfnamefont {Y.}~\bibnamefont {Katsuyama}},\ and\ \bibinfo {author} {\bibfnamefont {Y.}~\bibnamefont {Osada}},\ }\bibfield  {title} {\bibinfo {title} {Friction of gels},\ }\href@noop {} {\bibfield  {journal} {\bibinfo  {journal} {The Journal of Physical Chemistry B}\ }\textbf {\bibinfo {volume} {101}},\ \bibinfo {pages} {5487} (\bibinfo {year} {1997})}\BibitemShut {NoStop}%
\bibitem [{\citenamefont {Wu}\ \emph {et~al.}(2021)\citenamefont {Wu}, \citenamefont {Harper},\ and\ \citenamefont {Burton}}]{wu2021relaxation}%
  \BibitemOpen
  \bibfield  {author} {\bibinfo {author} {\bibfnamefont {B.}~\bibnamefont {Wu}}, \bibinfo {author} {\bibfnamefont {J.~M.}\ \bibnamefont {Harper}},\ and\ \bibinfo {author} {\bibfnamefont {J.~C.}\ \bibnamefont {Burton}},\ }\bibfield  {title} {\bibinfo {title} {Relaxation and recovery in hydrogel friction on smooth surfaces},\ }\href@noop {} {\bibfield  {journal} {\bibinfo  {journal} {Experimental Mechanics}\ }\textbf {\bibinfo {volume} {61}},\ \bibinfo {pages} {1081} (\bibinfo {year} {2021})}\BibitemShut {NoStop}%
\bibitem [{\citenamefont {Cuccia}\ \emph {et~al.}(2020)\citenamefont {Cuccia}, \citenamefont {Pothineni}, \citenamefont {Wu}, \citenamefont {M{\'e}ndez~Harper},\ and\ \citenamefont {Burton}}]{Cuccia2020}%
  \BibitemOpen
  \bibfield  {author} {\bibinfo {author} {\bibfnamefont {N.~L.}\ \bibnamefont {Cuccia}}, \bibinfo {author} {\bibfnamefont {S.}~\bibnamefont {Pothineni}}, \bibinfo {author} {\bibfnamefont {B.}~\bibnamefont {Wu}}, \bibinfo {author} {\bibfnamefont {J.}~\bibnamefont {M{\'e}ndez~Harper}},\ and\ \bibinfo {author} {\bibfnamefont {J.~C.}\ \bibnamefont {Burton}},\ }\bibfield  {title} {\bibinfo {title} {Pore-size dependence and slow relaxation of hydrogel friction on smooth surfaces},\ }\href@noop {} {\bibfield  {journal} {\bibinfo  {journal} {Proc. Natl. Acad. Sci. U.S.A.}\ }\textbf {\bibinfo {volume} {117}},\ \bibinfo {pages} {11247} (\bibinfo {year} {2020})}\BibitemShut {NoStop}%
\bibitem [{\citenamefont {Bico}\ \emph {et~al.}(2018)\citenamefont {Bico}, \citenamefont {Reyssat},\ and\ \citenamefont {Roman}}]{Bico2018}%
  \BibitemOpen
  \bibfield  {author} {\bibinfo {author} {\bibfnamefont {J.}~\bibnamefont {Bico}}, \bibinfo {author} {\bibfnamefont {{\'E}.}~\bibnamefont {Reyssat}},\ and\ \bibinfo {author} {\bibfnamefont {B.}~\bibnamefont {Roman}},\ }\bibfield  {title} {\bibinfo {title} {Elastocapillarity: When surface tension deforms elastic solids},\ }\href@noop {} {\bibfield  {journal} {\bibinfo  {journal} {Annual Review of Fluid Mechanics}\ }\textbf {\bibinfo {volume} {50}},\ \bibinfo {pages} {629} (\bibinfo {year} {2018})}\BibitemShut {NoStop}%
\bibitem [{\citenamefont {Davidovitch}\ \emph {et~al.}(2019)\citenamefont {Davidovitch}, \citenamefont {Sun},\ and\ \citenamefont {Grason}}]{Davidovitch2019}%
  \BibitemOpen
  \bibfield  {author} {\bibinfo {author} {\bibfnamefont {B.}~\bibnamefont {Davidovitch}}, \bibinfo {author} {\bibfnamefont {Y.}~\bibnamefont {Sun}},\ and\ \bibinfo {author} {\bibfnamefont {G.~M.}\ \bibnamefont {Grason}},\ }\bibfield  {title} {\bibinfo {title} {Geometrically incompatible confinement of solids},\ }\href@noop {} {\bibfield  {journal} {\bibinfo  {journal} {Proc. Natl. Acad. Sci. U.S.A.}\ }\textbf {\bibinfo {volume} {116}},\ \bibinfo {pages} {1483} (\bibinfo {year} {2019})}\BibitemShut {NoStop}%
\bibitem [{\citenamefont {King}\ \emph {et~al.}(2012)\citenamefont {King}, \citenamefont {Schroll}, \citenamefont {Davidovitch},\ and\ \citenamefont {Menon}}]{King2012}%
  \BibitemOpen
  \bibfield  {author} {\bibinfo {author} {\bibfnamefont {H.}~\bibnamefont {King}}, \bibinfo {author} {\bibfnamefont {R.~D.}\ \bibnamefont {Schroll}}, \bibinfo {author} {\bibfnamefont {B.}~\bibnamefont {Davidovitch}},\ and\ \bibinfo {author} {\bibfnamefont {N.}~\bibnamefont {Menon}},\ }\bibfield  {title} {\bibinfo {title} {Elastic sheet on a liquid drop reveals wrinkling and crumpling as distinct symmetry-breaking instabilities},\ }\href@noop {} {\bibfield  {journal} {\bibinfo  {journal} {Proc. Natl. Acad. Sci. U.S.A.}\ }\textbf {\bibinfo {volume} {109}},\ \bibinfo {pages} {9716} (\bibinfo {year} {2012})}\BibitemShut {NoStop}%
\bibitem [{\citenamefont {Matan}\ \emph {et~al.}(2002)\citenamefont {Matan}, \citenamefont {Williams}, \citenamefont {Witten},\ and\ \citenamefont {Nagel}}]{Matan2002}%
  \BibitemOpen
  \bibfield  {author} {\bibinfo {author} {\bibfnamefont {K.}~\bibnamefont {Matan}}, \bibinfo {author} {\bibfnamefont {R.~B.}\ \bibnamefont {Williams}}, \bibinfo {author} {\bibfnamefont {T.~A.}\ \bibnamefont {Witten}},\ and\ \bibinfo {author} {\bibfnamefont {S.~R.}\ \bibnamefont {Nagel}},\ }\bibfield  {title} {\bibinfo {title} {Crumpling a thin sheet},\ }\href@noop {} {\bibfield  {journal} {\bibinfo  {journal} {Phys. Rev. Lett.}\ }\textbf {\bibinfo {volume} {88}},\ \bibinfo {pages} {076101} (\bibinfo {year} {2002})}\BibitemShut {NoStop}%
\bibitem [{\citenamefont {Lomholt}\ \emph {et~al.}(2013)\citenamefont {Lomholt}, \citenamefont {Lizana}, \citenamefont {Metzler},\ and\ \citenamefont {Ambj\"ornsson}}]{Lomholt2013}%
  \BibitemOpen
  \bibfield  {author} {\bibinfo {author} {\bibfnamefont {M.~A.}\ \bibnamefont {Lomholt}}, \bibinfo {author} {\bibfnamefont {L.}~\bibnamefont {Lizana}}, \bibinfo {author} {\bibfnamefont {R.}~\bibnamefont {Metzler}},\ and\ \bibinfo {author} {\bibfnamefont {T.}~\bibnamefont {Ambj\"ornsson}},\ }\bibfield  {title} {\bibinfo {title} {Microscopic origin of the logarithmic time evolution of aging processes in complex systems},\ }\href@noop {} {\bibfield  {journal} {\bibinfo  {journal} {Phys. Rev. Lett.}\ }\textbf {\bibinfo {volume} {110}},\ \bibinfo {pages} {208301} (\bibinfo {year} {2013})}\BibitemShut {NoStop}%
\end{thebibliography}%

\end{document}



\title{Supplementary Information:\\Geometry-induced friction at a soft interface}

\author{Aashna Chawla}
\author{Deepak Kumar}%
 \email{krdeepak@physics.iitd.ac.in}
\affiliation{Department of Physics, Indian Institute of Technology Delhi, Hauz Khas, New Delhi, India
}%

\maketitle


\subsection{Slope $m$ as a measure of friction} \label{S1}

We fit a straight line to the initial part $(1>\tilde{R}\gtrsim 0.9)$ of the $\tilde{W}(\tilde{R})$ curve to obtain the slope $m$. The slope $m$ is related to the relative velocity between the sheet and the substrate and is a measure of the strength of frictional interaction between the sheet and the substrate in the same way as in a sedimentation experiment the terminal velocity can be used to measure the drag coefficient. In this section, we develop the relationship between $m$ and the strength of frictional interaction.

\subsubsection{Relative velocity}
For simplicity, we consider the planar geometry shown in Fig. \ref{fig:relvel}. A sheet with initial radius $W_0$ is compressed isotropically such that its radius changes by a factor $\tilde{W}=\frac{W}{W_0}$ after a time interval $t$. In this process, a material point at $(r,\theta)$ goes to $(r',\theta)$, such that $r'=r\frac{W}{W_0}$. The radial displacement $\Delta r=r'-r=r\frac{\Delta W}{W_0}$ and the radial velocity $u(r)=\frac{dr}{dt}=r\frac{d\tilde{W}}{dt}$. Using expressions for the velocities of the sheet ($u_{sheet}$) and the substrate ($u_{sub}$), the relative velocity between the two can be written as:

\begin{equation}
\label{eqn:1}
    v(r)= u_{sub}(r)-u_{sheet}(r)\\
    =r \left( \frac{d\tilde{R}}{dt}-\frac{d\tilde{W}}{dt}\right)=r\frac{d\tilde{R}}{dt}(1-m)   
\end{equation}

\subsubsection{Balance of forces}
Based on previous studies that have found the friction in hydrogels to be velocity dependent: $F_{fr}\propto v^{\alpha}$, with $\alpha\sim1$ \cite{Simi2020,gong1997friction,wu2021relaxation,Cuccia2020}, we assume a frictional interaction of the form $F_{fr}=\Gamma v$, where $v=W_0\frac{d\tilde{R}}{dt}(1-m)$ is the relative velocity between the sheet and the substrate and $\Gamma$ represents strength of the frictional interaction. The frictional force $F_{fr}$ tends to compress the sheet. This effect is opposed by a restoring force of the form $F_{el}=YdW$, where $Y$ is the stretching stiffness of the sheet. The steady state is characterized by a balance of these two forces:

\begin{figure}[!htb]
    \begin{center}
        \includegraphics[scale=0.2]{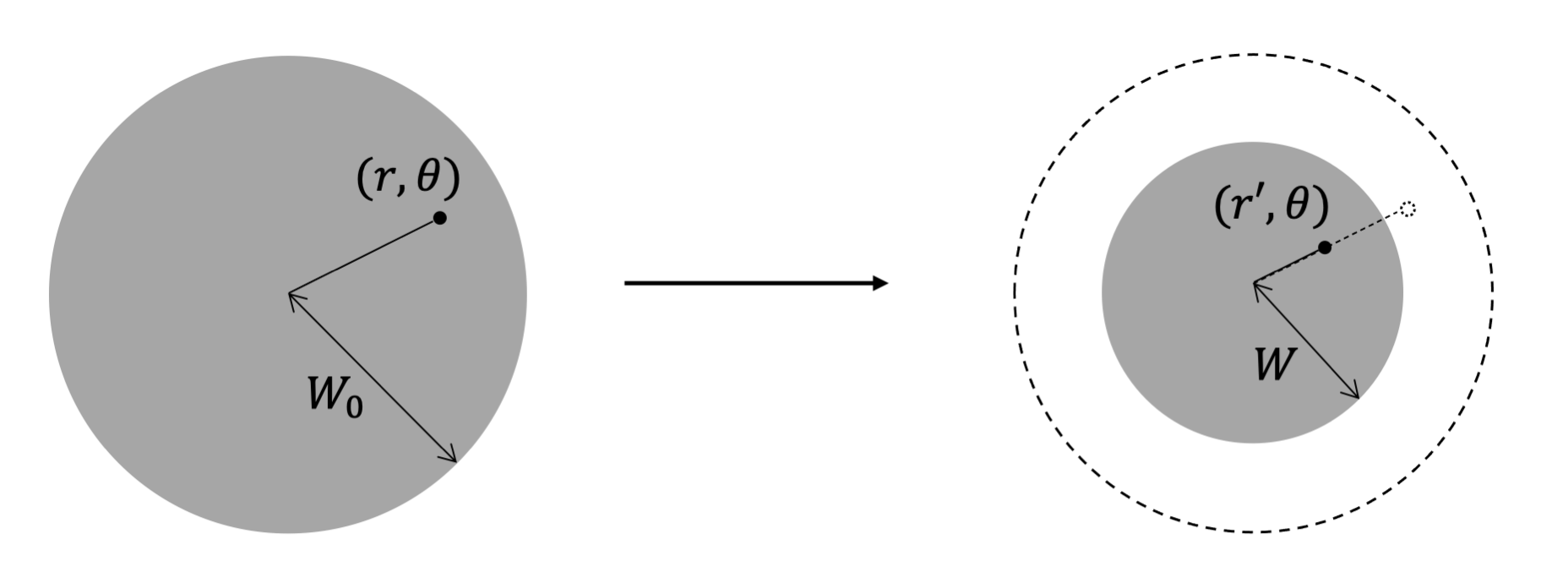}
    \end{center}
    \caption{A flat sheet of radius $W_0$ is compressed isotropically to radius $W$. In this process, the material point at $(r,\theta)$ is displaced to the point $(r',\theta)$.}
   \label{fig:relvel}
\end{figure}

\begin{equation}
\label{eqn:2}
\Gamma v= Y dW   
\end{equation}

\begin{equation}
    \implies \Gamma W_0\frac{d\tilde{R}}{dt}(1-m)=Y dW
\end{equation}

\begin{equation}
    \implies \Gamma= \frac{m}{1-m}Y dt
\end{equation}
where we have used the fact that $m=d\Tilde{W}/d\Tilde{R}$. For $m\ll1$, we get:
\begin{equation}
    \Gamma \propto m
\end{equation}

Thus, $m$ is a measure of the strength of frictional interaction between the sheet and the substrate.

\subsection{Relative contributions of various forces to the energy density and pressure}
We consider a thin flat sheet of radius $W$, thickness $h$, and Young's modulus $E$ placed on a sphere of radius $R$ and surface energy $\gamma$ and estimate the order of magnitude contributions of stretching, bending, surface energy and gravity to the energy density (per unit area) of the sheet and the normal pressure at the interface. We use the following typical values of the parameters: $W\sim 1mm$, $h=100nm$, $E=3GPa$, $\gamma \sim 0.01mN/m$ and $g=10m/s^2$. Table \ref{tab:comparison} lists scaling relations \cite{Bico2018,Davidovitch2019,King2012} and the estimates of the order of magnitude contributions of the various terms.

Considering the values listed in Table \ref{tab:comparison}, we note that contributions from both bending and gravity are much smaller than that from strain in the sheet and due to surface energy. 
Further, we note that for the values of parameters considered, pressure due to geometrical strain is much larger than the pressure due to surface energy, and therefore, geometrical strain has a much larger effect on friction. While surface energy is important in the adhesion of the flat sheet to the spherical substrate, friction at the interface is dominated by contribution from strain in the sheet due to its geometrically incompatible confinement to the spherical substrate.

\begin{table}\centering
\renewcommand{\arraystretch}{2}
    \centering
    \begin{tabular}{| c | c | c |}
    \hline
       \textbf{Contribution} & \textbf{Energy density [N/m]} &\textbf {Pressure [Pa]}\\
         \hline
         Stretching & $Y (\frac{W}{R})^4 \sim 10^{-1} $ &$ \frac{Y}{R}(\frac{W}{R})^2 \sim 10^3 $\\
         \hline 
         Surface energy & $ \gamma\sim 0.01$ & $ \frac{\gamma}{R}\sim 1$\\ 
         \hline
         Bending & $ \frac{B}{R^2}\sim 10^{-9}$ & $\frac{B}{R^3}\sim 10^{-7}$\\
         \hline
         Gravity & $\rho g h R \sim 10^{-6}$ & $\rho g h \sim 10^{-3}$\\ 
         \hline
    \end{tabular}
    \caption{Order of magnitude value for the contribution of various forces to energy density and pressure.}
    \label{tab:comparison}
\end{table}

\subsection{Additional details about cylindrical substrate}
\begin{figure}
    \begin{center}
        \includegraphics[width=0.75\textwidth]{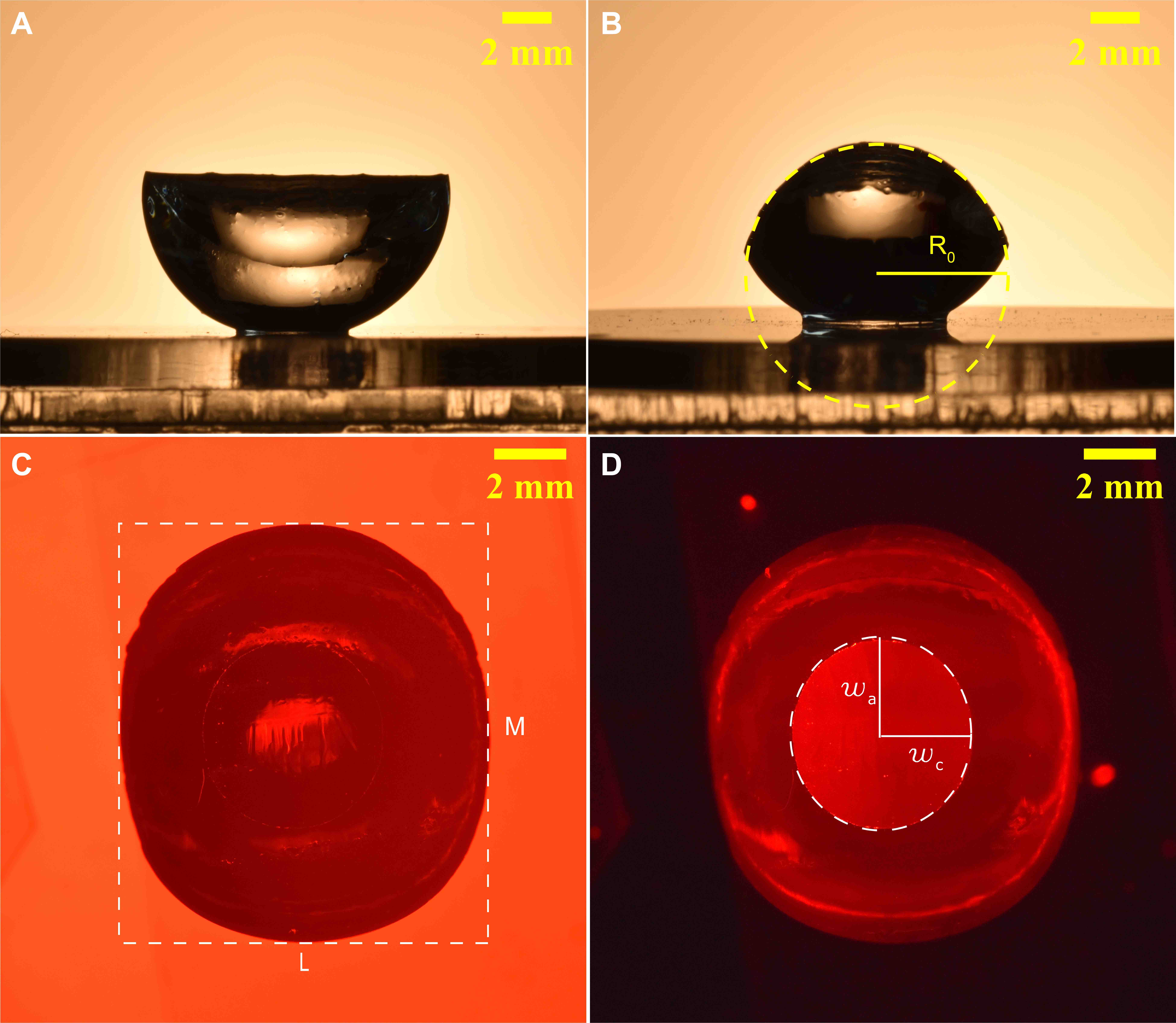}
    \end{center}
    \caption{Experiments on cylindrical substrate: (A) Side-view image of the cylindrical substrate along the axial direction. (B) Side-view image of the cylindrical substrate along the curved direction. The dashed yellow curve shows the circle with the measured radius. (C) A typical image of the cylindrical substrate during the experiment. White dashed curve shows a rectangle of dimensions $L \times M$ that best fits the substrate. (D) A typical image of the sheet on the cylindrical substrate. White dashed curve shows the best-fit ellipse with the semi-axis lengths $w_a$ and $w_c$ measured along the axial and curved directions, respectively.}
      \label{fig:cylinder}
\end{figure}

We obtain cylindrical hydrogel substrate by cutting a swollen hydrogel sphere using a thin sharp blade bent in the form of a cylinder. Figure \ref{fig:cylinder}A,B show two cross-sectional side views of a typical cylindrical surface so obtained - first one (Fig. \ref{fig:cylinder}A) is along the axial direction and second one (Fig. \ref{fig:cylinder}B)  is along the curved direction.  Figure \ref{fig:cylinder}B is used to determine initial diameter $2R_0$ of the cylinder. 

Figure \ref{fig:cylinder}C,D show the typical top view images acquired in experiments with cylindrical substrates using the method described in the main text. Brightfield images of the substrate, such as Fig. \ref{fig:cylinder}C, are analyzed to determine the best fit bounding rectangle (with edge lengths $L\times M$), as illustrated in the figure by dashed white lines. Assuming that the evaporation-induced shrinking of the hydrogel is isotropic, we determine the diameter of the cylinder at this instant as $2R = 2R_0 \left(\frac{L}{L_0}\right)$, where $L_0$ is the value of $L$ at the beginning of the experiment.

Fluorescent images such as Fig. \ref{fig:cylinder}D are analyzed to obtain the sheet radius at any instant of time. We find the ellipse (with semi-major and semi-minor axes lengths $w_a$ and $w_c$, respectively) that best fits the sheet image. The semi-major axis length measures $W_a=w_a$, the sheet radius along the axial direction. The radius of the sheet along the curved direction is determined from the semi-minor axis length $w_c$ as: $W_c=R sin^{-1}(w_c/R)$. Variations of the normalized sheet radii along the axial direction, $\tilde{W_a}=W_a/W_{a0}$, and along the curved direction, $\tilde{W_c}=W_c/W_{c0}$, with the normalized length $\tilde{M}=M/M_0$ and normalized radius of the cylinder $\tilde{R}=R/R_0$ are used to determine the slopes $m_a$ and $m_c$, respectively. 
\subsection{Additional details about flat substrate}
\begin{figure}[!h]
    \begin{center}
        \includegraphics[width=0.75\textwidth]{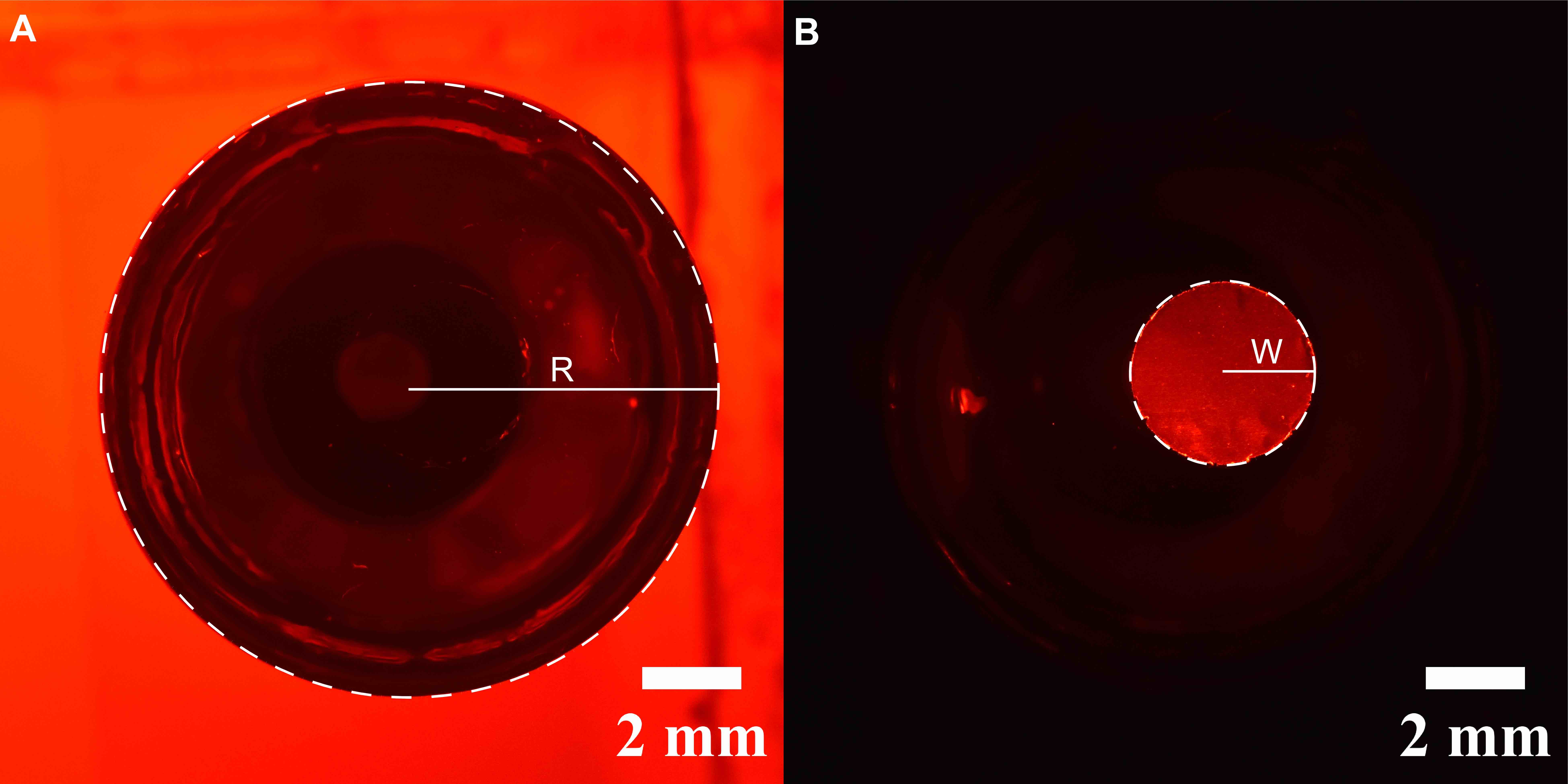}
    \end{center}
    \caption{Flat sheet on a flat substrate. (A) A typical image of the flat hydrogel substrate (with the sheet on the top) captured during the experiment. White dashed curve shows the best-fit ellipse, with semi-major axis length $R$ used to measure the substrate size. (B) A typical image of the sheet. The white dashed curve shows a circle of radius $W$ obtained by the radial profile method described in the methods section.}
   \label{fig:flat}
\end{figure}

Flat hydrogel substrates are obtained by cutting a swollen hydrogel sphere into two halves using a thin, flat blade. A flat sheet is then placed on the planar hydrogel surface. Figures \ref{fig:flat}A,B show the typical images of the hydrogel substrate and the sheet obtained during the experiment. The bright-field image of the hydrogel substrate is analyzed to determine the best-fit ellipse, shown with a white dashed line. The semi-major axis of the best-fit ellipse is taken as the measure of the substrate size $R$. The fluorescent image of the sheet is analyzed to determine its radius $W$.


\subsection{Note on roughness of the substrates}
\begin{figure}[!h]
    \begin{center}
    \includegraphics[width=0.8\textwidth]{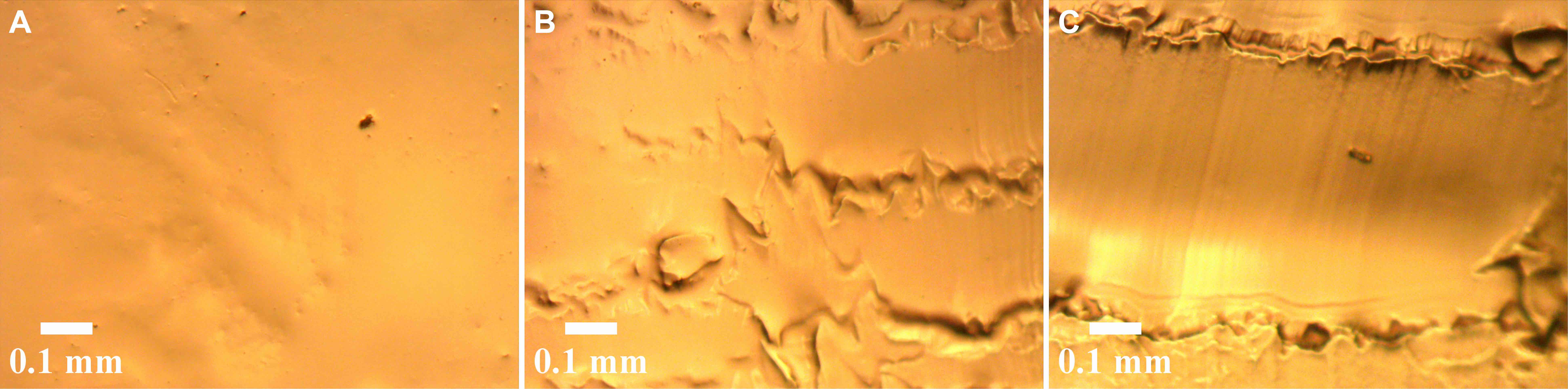}
    \end{center} 
    \caption{Typical optical micrographs of swollen hydrogel surfaces (A) Spherical hydrogel surface (B) Flat hydrogel surface (C) Cylindrical hydrogel surface. Flat and cylindrical surfaces appear rougher in comparison to the spherical surface. The process of cutting with a blade used to obtain the flat and cylindrical surfaces from hydrogel spheres introduces roughness in the surfaces.}
    \label{fig:roughness}
\end{figure}

Figure \ref{fig:roughness}A shows an optical micrograph of the surface of a swollen hydrogel sphere. We note that the surface is quite smooth. On the other hand, the process of cutting used to obtain flat and cylindrical substrates introduces some amount of roughness in the hydrogel surface. This can be seen in Fig. \ref{fig:roughness}B,C, where we have shown optical micrographs for the flat and cylindrical substrates, respectively.

\subsection{Note on computation of error bars}
The error bars in Fig. 2A,B represent the errors in calculating $m$ by fitting a straight line to the $\tilde{W}(\tilde{R})$ data. We have included in this calculation the effect of the error ($\sim$ 1 pixel) in measuring $W$ from the images. In Fig. 2C, in addition to the above sources of error, we have also included the effect of uncertainty in the measurement of the radius of curvature of the cylindrical substrate, which contributes to errors both on the $y$ (slope) and the $x$ (mean curvature) axes.



\begin{figure}
    \begin{center}
    \includegraphics[scale=0.75]{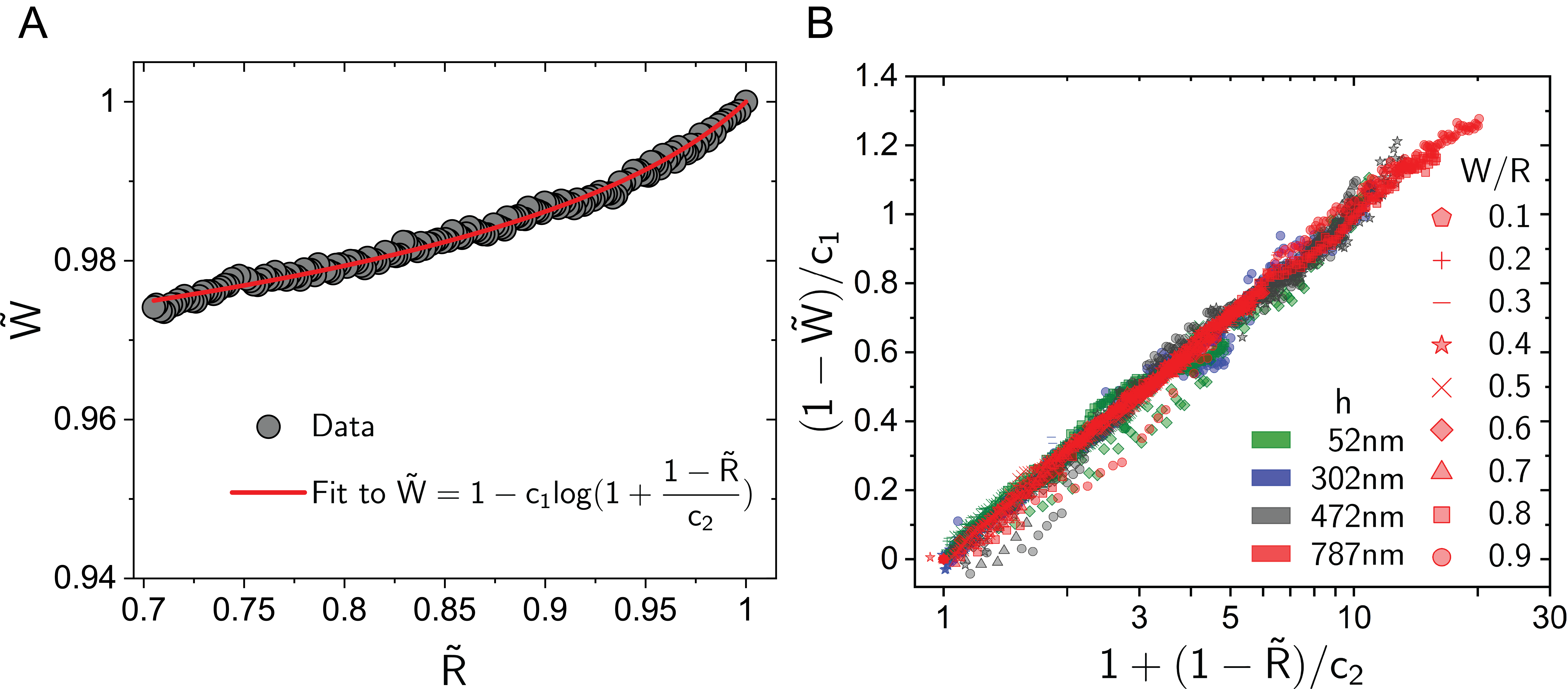}
    \end{center} 
    \caption{ \textbf{Logarithmic variation of $\Tilde{W}(\Tilde{R})$:} (A) A typical $\Tilde{W}(\Tilde{R})$ data (grey circles) plotted over $0.7<\tilde{R}<1$ along with a fit to the function $\Tilde{W}=1-c_1 \log(1+\frac{1-\Tilde{R}}{c_2})$, with $c_1$ and $c_2$ as fitting parameters. (B) Various data sets for sheets of thickness $h=52nm$, $302nm$, $472nm$, and $787nm$, and with $W_0/R_0$ ranging from $0.1$ to $0.9$ placed on spherical hydrogel substrates, plotted as $\frac{1-\tilde{W}}{c_1}$ \emph{versus} $1+\frac{1-\tilde{R}}{c_2}$ on a semi-log plot. All the data collapses to a single straight line when plotted in this manner.}

     \label{fig:log}
\end{figure}

\subsection{Logarithmic variation of $\tilde{W}$ with $\tilde{R}$}
In the main text, we have calculated the slope $m$ of the $\tilde{W}(\tilde{R})$ curve by fitting a straight line to a small region near $\tilde{R}=1$. In general, over a wider range of $\tilde{R}$, we find that the curve $\tilde{W}(\tilde{R})$ is nonlinear. Figure \ref{fig:log}A shows $\tilde{W}(\tilde{R})$ for a sheet of thickness $h=52nm$ and $W_0/R_0=0.6$. We find empirically that the following logarithmic function describes the $\tilde{W}(\tilde{R})$ curve well:

\begin{equation}
    \Tilde{W}=1-c_1 \log(1+\frac{1-\Tilde{R}}{c_2})
    \label{eq:logfunc1}
\end{equation},
where $c_1$ and $c_2$ are constants.

In Fig. \ref{fig:log}A, the red curve shows a fit of this function to the data. We fit the above functional form to the data obtained for various values of $h$ and $\frac{W_0}{R_0}$ and find the values of $c_1$ and $c_2$ in each case. In Fig. \ref{fig:log}B, we plot all these data together on a semi-log plot keeping $1+\frac{1-\Tilde{R}}{c_2}$ on the $x$-axis and $\frac{1-\Tilde{W}}{c_1}$ on the $y$-axis. On this log-linear scale, all these different data sets fall on a straight line, implying that the functional form describes the data well. The functional form is reminiscent of the logarithmic response observed in many experiments where the strain in a sheet that is allowed to crumple under a constant external load has been measured with time. In fact, we can rewrite Equation \ref{eq:logfunc1} as:

\begin{equation}
\frac{\epsilon}{c_1}=log(1+\frac{\dot{\Tilde{R}} t}{c_2})
\label{eq:logfunc2}
\end{equation}

 where $\epsilon=1-\Tilde{W}$ is the strain in the sheet and $\dot{\Tilde{R}}=\frac{d\Tilde{R}}{dt}$ is the rate of change of radius of the hydrogel sphere, which is observed to be almost constant in a given experiment. The form of Equation \ref {eq:logfunc2} is identical to the response of thin sheets observed in other crumpling experiments \cite{Matan2002, Lomholt2013}. This analogy suggests two important features of our system: force of friction is nearly constant over a large range of $\Tilde{R}$ and that the force resisting compression of the film is predominantly related to the elastic response of the sheet.

\subsection*{References}

\bibliography{ref2}